\documentclass[prd,aps,showpacs,tightenlines,floats,nofootinbib,preprintnumbers]{revtex4}

\usepackage{graphicx}
\usepackage{amsmath,amssymb,algorithm,algorithmic,color}

\newcommand{\pf}[2]{\frac{\partial#1}{\partial#2}}
\newcommand{\ppf}[2]{\frac{\partial^2#1}{\partial#2^2}}

\begin{document}

\title{Instability of colliding metastable strings}

\author{
Takashi Hiramatsu$^1$, 
Minoru Eto$^2$, 
Kohei Kamada$^3$\footnote{
Present address:  Institut de Th\'eorie des Ph\'enom\`enes Physiques,
\'Ecole Polytechnique F\'ed\'erale de Lausanne,
CH-1015 Lausanne, Switzerland
}, 
Tatsuo Kobayashi$^4$ 
and Yutaka Ookouchi$^{4,5}$
}

\affiliation{
$^1$ Yukawa Institute for Theoretical Physics, Kyoto University, Kyoto 606-8502, Japan \\ 
$^2$ Department of Physics, Yamagata University, Yamagata 990-8560, Japan \\
$^3$ Deutsches Elektronen-Synchrotron DESY, Notkestra\ss e 85, D-22607 Hamburg, Germany \\
$^4$ Department of Physics, Kyoto University, Kyoto 606-8502, Japan \\
$^5$The Hakubi Center for Advanced Research, Kyoto University, Kyoto 606-8302, Japan
} 

\preprint{YITP-13-19,~YGHP-12-55,~DESY~13-039,~KUNS-2439}

%%%%%%%%%%%%%%%%%%%%%%%%%%%%%%%%%%%%%%%%%%%%%%%%%%%%%%%%%%%%%%%%%%%%%%%%%%%%%%%
\begin{abstract}
The breaking of $U(1)_R$ symmetry plays a crucial role in modeling the
breaking of supersymmetry (SUSY). In the models that possess both SUSY
preserving and SUSY breaking vacua, tube-like cosmic strings called
R-tubes, whose surfaces are constituted by domain walls interpolating a
false and a true vacuum with some winding numbers, can exist. Their
(in)stability can strongly constrain SUSY breaking models
theirselves. In the present study, we investigate the dynamical
(in)stability of two colliding metastable tube-like strings by
field-theoretic simulations. From them, we find that the strings become
unstable, depending on the relative collision angle and speed of two
strings, and the false vacuum is eventually filled out by the true vacuum
owing to rapid expansion of the strings or unstable bubbles created as
remnants of the collision. 
\end{abstract}
%%%%%%%%%%%%%%%%%%%%%%%%%%%%%%%%%%%%%%%%%%%%%%%%%%%%%%%%%%%%%%%%%%%%%%%%%%%%%%%
\pacs{11.27.+d, 98.80.Cq, 98.80.-k}

\maketitle

%%%%%%%%%%%%%%%%%%%%%%%%%%%%%%%%%%%%%%%%%%%%%%%%%%%%%%%%%%%%%%%%%%%%%%%%%%%%%%%
%%%%%%%%%%%%%%%%%%%%%%%%%%%%%%%%%%%%%%%%%%%%%%%%%%%%%%%%%%%%%%%%%%%%%%%%%%%%%%%
\section{introduction}
%%%%%%%%%%%%%%%%%%%%%%%%%%%%%%%%%%%%%%%%%%%%%%%%%%%%%%%%%%%%%%%%%%%%%%%%%%%%%%%
%%%%%%%%%%%%%%%%%%%%%%%%%%%%%%%%%%%%%%%%%%%%%%%%%%%%%%%%%%%%%%%%%%%%%%%%%%%%%%%

Global $U(1)$ symmetry breaking plays a crucial role in many aspects of particle physics and field theories. 
Cosmic strings are generated associated with the $U(1)$ symmetry breaking in the cosmic history and 
hence are the key solitons to probe information of the 
early universe or to prove (or disprove) a phenomenological model
\cite{Vilenkin}. 

One of the most important global $U(1)$ symmetries is the $U(1)_R$ symmetry in supersymmetric theories. 
Since supersymmetry (SUSY) is broken at low-energy scales, we must live in a vacuum with 
nonvanishing potential energy, which requires exact $U(1)_R$ symmetry if the vacuum is the 
global minimum. On the other hand, exact $U(1)_R$ symmetry prohibits nonzero Majorana gaugino mass. 
In this reason,   relatively complicated vacuum structure that contains both true SUSY vacua and false vacua 
with approximate $U(1)_R$ symmetry (broken explicitly or spontaneously) is now energetically studied for 
low-energy model building in the phenomenological point of view.  (See \cite{SUSY1,SUSY2} for reviews.) 
%Recent developments of the study of the string landscape \cite{landscape1,landscape2} strongly supports this idea 
%since it suggests the vacuum structure in the string theory, which is the most promising candidate of the 
%quantum gravity, is quite involved. 
%Like the above example, in a wide class of models, 
When a false vacuum is selected in such models after inflation, the
$U(1)_R$ symmetry is spontaneously broken and it gives rise to cosmic
R-strings, which cannot be avoided in a high scale inflation or a high
reheating temperature scenario. 

Cosmic strings in some models of this class have a peculiar feature. 
If a lower $U(1)$ preserving vacuum exists, 
the field configuration in the string core falls down to the lower vacuum. Therefore, the strings have inner structure that
is tube-like domain wall configuration.  
Note that since the lower vacuum is energetically favored, 
the interior of the string core possibly starts to expand
\cite{Steinhardt:1981ec,Hosotani,Yajnik:1986tg,Eto:2012ij,Kamada:2013rya},
and the Universe is eventually filled by the true vacuum, where
SUSY is recovered. This situation is inconsistent with our Universe.

Besides this extreme case, it has been also reported that there is
parameter space where the meta-stable tube-like strings exist
\cite{Eto:2012ij}. 
However, naively thinking, such an object would be unstable
against any disturbance, and follow the same fate as the above case.
Actually, once multiple strings are formed by the breaking of $U(1)_R$
in the cosmic history, they are
inevitably affected by the violent dynamical processes like collision and
reconnection. Taking such kinds of dynamical processes of meta-stable
R-strings into account, it is non-trivial whether the tube-like strings become
unstable or keep their stability throughout their collision process. 
Thus, it is very important to study the collision dynamics and figure
out the (in)stability of the tube-like R-strings, which has a tight
relationship with the possibility to constrain the model building for
realistic SUSY breaking and $U(1)_R$ breaking.

In this paper, we investigate the dynamical stability/instability for
the strings under the reconnection. For the sake of this study, we
perform the field-theoretic simulations of the collision of two meta-stable
strings described as a solitonic solution in the model of a classical complex
scalar field with a potential including a true vacuum and a meta-stable
vacuum. 
For the simulations, we particularly focus on the two physical
parameters characterising the collision process; the relative velocity
and the angle of two colliding strings. 
As a result, we find that the stability of the system strongly
depends on these parameters and that surprisingly there is a wide parameter space
where the system is stable against the collision.
This work implies the importance of taking into account the
dynamics of the meta-stable objects generated in the SUSY breaking and
$U(1)_R$ breaking models.
%Here, we focus on a simple single field model with a false vacuum and
%a true vacuum. 
%This is partially because the reconnection process is
%highly involved and is not easy to analyze complicated models with two
%or more fields by computer simulations in reliable precision. 
Although this system cannot produce the cosmic string network via the Kibble mechanism, 
the process of the string collision would catch up the feature of more realistic situations. 
At any rate, it is very useful to reveal  novel phenomena by exploiting a simple model 
which would be shared by a wide class of theories. 
 
The organization of this paper is as follows. In section II, we set up
the model and show numerical solutions for static metastable strings. In
section III, by using approximations, we estimate the maximum winding
number of a metastable string and study analytically instability of
colliding two strings. In section IV, we investigate the dynamics of
colliding strings by three-dimensional simulations. We survey parameter dependence of
instability by varying the collision angle and the relative speed
of strings. Section V is devoted to conclusions and discussions. 
In Appendix, we briefly explain our scheme for numerical studies.

%%%%%%%%%%%%%%%%%%%%%%%%%%%%%%%%%%%%%%%%%%%%%%%%%%%%%%%%%%%%%%%%%%%%%%%%%%%%%%%
%%%%%%%%%%%%%%%%%%%%%%%%%%%%%%%%%%%%%%%%%%%%%%%%%%%%%%%%%%%%%%%%%%%%%%%%%%%%%%%
\section{set-up of model and global string}
\label{sec:model}
%%%%%%%%%%%%%%%%%%%%%%%%%%%%%%%%%%%%%%%%%%%%%%%%%%%%%%%%%%%%%%%%%%%%%%%%%%%%%%%
%%%%%%%%%%%%%%%%%%%%%%%%%%%%%%%%%%%%%%%%%%%%%%%%%%%%%%%%%%%%%%%%%%%%%%%%%%%%%%%

\begin{figure}[htbp]
\centering{
 \includegraphics[width=.45\linewidth,bb=0 0 259 160]{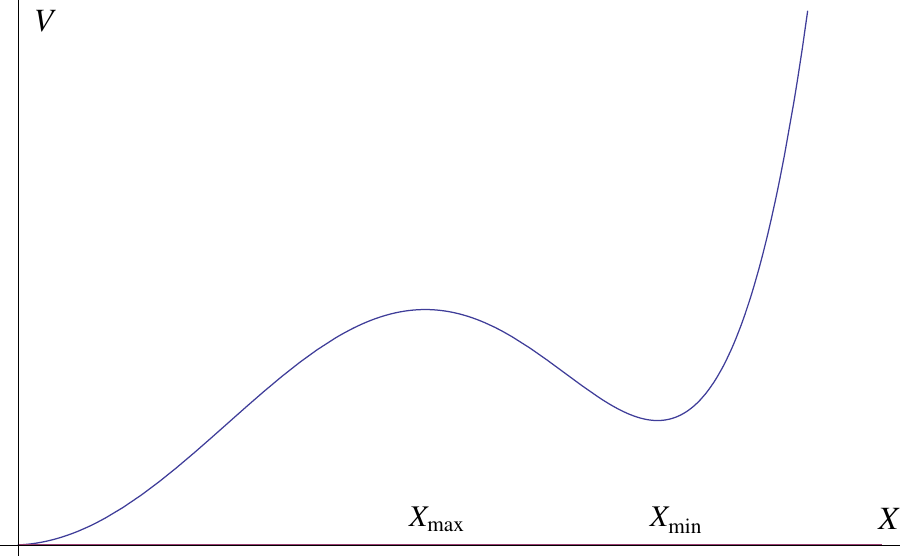}
}
\caption{The sixth order of potential Eq.~\eqref{eq:sixthpotential}. On the
 false vacuum at $X=X_{min}$, the global $U(1)$ symmetry is
 spontaneously broken.} 
\label{fig:potentialshape}
\end{figure}

To illustrate growing instability of metastable strings under
reconnections and show generality of such phenomena, we consider a
simple single field model with false and true vacua. This is an ideal
example to demonstrate various aspects of the collision dynamics of
two metastable solitons which would be common in a wide class of
models. The Lagrangian of a complex scalar field $X$ which carries a
charge of global $U(1)$ symmetry is given by 
%%%%%%%%%%%%%%%%%%%%%%%%%%%%%%%%%%%%%%%%%%%%%%%%%%%%%%%%%%%%%%%%%%%%
%
\begin{equation}
{\cal L}=|\partial_{\mu} X|^2+V(X).
\end{equation}
%
%%%%%%%%%%%%%%%%%%%%%%%%%%%%%%%%%%%%%%%%%%%%%%%%%%%%%%%%%%%%%%%%%%%%
We engineer a false vacuum and a true vacuum by employing the following
sixth order of the potential,
%%%%%%%%%%%%%%%%%%%%%%%%%%%%%%%%%%%%%%%%%%%%%%%%%%%%%%%%%%%%%%%%%%%%
%
\begin{equation}
 V(X) = \mu^2 |X|^2 \left( \delta + {1\over M^{4}} ( |X|^2 - \eta^2 )^2\right), 
  \label{eq:sixthpotential}
\end{equation}
%
%%%%%%%%%%%%%%%%%%%%%%%%%%%%%%%%%%%%%%%%%%%%%%%%%%%%%%%%%%%%%%%%%%%%
where $\delta$ is a dimensionless constant determining 
the amplitude of a local minimum of $V(X)$ at $X \ne 0$. 
This potential has the global minimum at $X=0$ where $V=0$, a local
maximum at $X=X_{\rm max}$, and a local minimum at $X=X_{\rm min}$,
which are given by
%%%%%%%%%%%%%%%%%%%%%%%%%%%%%%%%%%%%%%%%%%%%%%%%%%%%%%%%%%%%%%%%%%%%
%
\begin{equation}
 \begin{aligned}
V_{\rm min} =  V(X_{\rm min}) = \eta^4\frac{2\epsilon^2\zeta^4}{27}(c+2)^2(1-c), &\quad
V_{\rm max} = V(X_{\rm max}) = \eta^4\frac{2\epsilon^2\zeta^4}{27}(c-2)^2(c+1), \\
|X_{\rm min}| = \eta\sqrt{\frac{2+c}{3}}, &\quad
|X_{\rm max}| = \eta\sqrt{\frac{2-c}{3}}. \label{eq:Xmaxmin}
 \end{aligned}
\end{equation}
%
%%%%%%%%%%%%%%%%%%%%%%%%%%%%%%%%%%%%%%%%%%%%%%%%%%%%%%%%%%%%%%%%%%%%
Here we introduced several dimensionless quantities,
$c=\sqrt{1-3\delta/\zeta^4}$, $\zeta = \eta/M$ and 
$\epsilon = \mu/\eta$.  Figure \ref{fig:potentialshape} shows the schematic
picture of this potential. The Euler-Lagrange equation we solve in
section IV for simulations of colliding strings is given by 
%%%%%%%%%%%%%%%%%%%%%%%%%%%%%%%%%%%%%%%%%%%%%%%%%%%%%%%%%%%%%%%%%%%%
%
\begin{equation}
\frac{\partial^2X}{\partial t^2}- \triangle X + \frac{dV}{dX^*} = 0.
 \label{eq:fieldeq}
\end{equation}
%
%%%%%%%%%%%%%%%%%%%%%%%%%%%%%%%%%%%%%%%%%%%%%%%%%%%%%%%%%%%%%%%%%%%%

%%%%%%%%%%%%%%%%%%%%%%%%%%%%%%%%%%%%%%%%%%%%%%%%%%
%%%%%%%%%%%%%%%%%%%%%%%%%%%%%%%%%%%%%%%%%%%%%%%%%%

In the false vacuum, the global $U(1)$ symmetry is spontaneously
broken. We simply assume that a global string is formed. Consider a
static cylindrical solution of the field equation in 
the cylindrical coordinate, $(r,\theta,z)$. 
First we decompose $X$ into the radial and angular parts, 
%%%%%%%%%%%%%%%%%%%%%%%%%%%%%%%%%%%%%%%%%%%%%%%%%%%%%%%%%%%%%%%%%%%%
%
\begin{equation}
  X = \eta R(r)e^{in\theta} \quad (n=1,2,3,\cdots).
\end{equation}
%
%%%%%%%%%%%%%%%%%%%%%%%%%%%%%%%%%%%%%%%%%%%%%%%%%%%%%%%%%%%%%%%%%%%%
Then the explicit form of the equation of $R(r)$ is obtained from Eq.~(\ref{eq:fieldeq})
%%%%%%%%%%%%%%%%%%%%%%%%%%%%%%%%%%%%%%%%%%%%%%%%%%%%%%%%%%%%%%%%%%%%
%
\begin{equation}
 \ppf{R}{x} + \frac{1}{x}\pf{R}{x}
 - \frac{n^2}{x^2}R
 = (\delta+\zeta^4)R + \zeta^4( 3R^5 - 4R^3),
 \label{eq:R}
\end{equation}
%
%%%%%%%%%%%%%%%%%%%%%%%%%%%%%%%%%%%%%%%%%%%%%%%%%%%%%%%%%%%%%%%%%%%%
where $x = \epsilon r\eta$. We look for the solution where
the scalar field in the interior of the string stays in the true vacuum,
and that stays in the meta-stable vacuum at the exterior. That is,
such a solution should satisfy $R=0$ at $x=0$ and $R\to R_{\rm min}$ at
$x\to \infty$.

Here we emphasize that there can be a (meta)stable tube-like solution. 
Since the string core has lower energy density than that of its exterior, 
larger string radius seems to be favorable for the system, which predicts 
the roll-over process. However, there must be the domain wall between
its core and exterior, whose energy becomes larger for larger string
radius. As a result, there arises a (meta)stable tube-like solution,
depending on the parameters. We will see it in more detail in the next
section.  

\begin{figure}[!ht]
\centering{
  \includegraphics[width=8.5cm,bb=0 0 792 612]{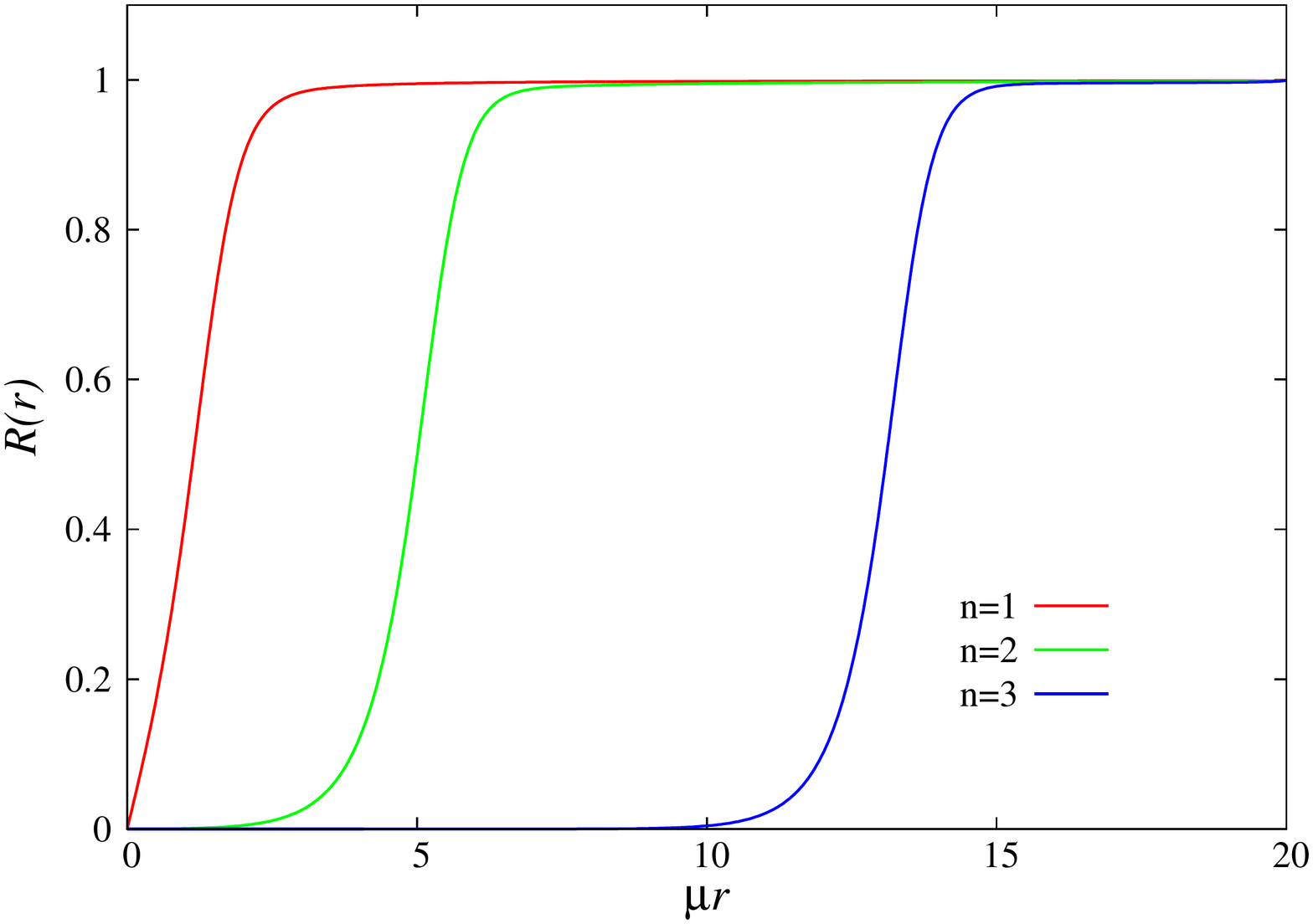}
  \includegraphics[width=8.5cm,bb=0 0 792 612]{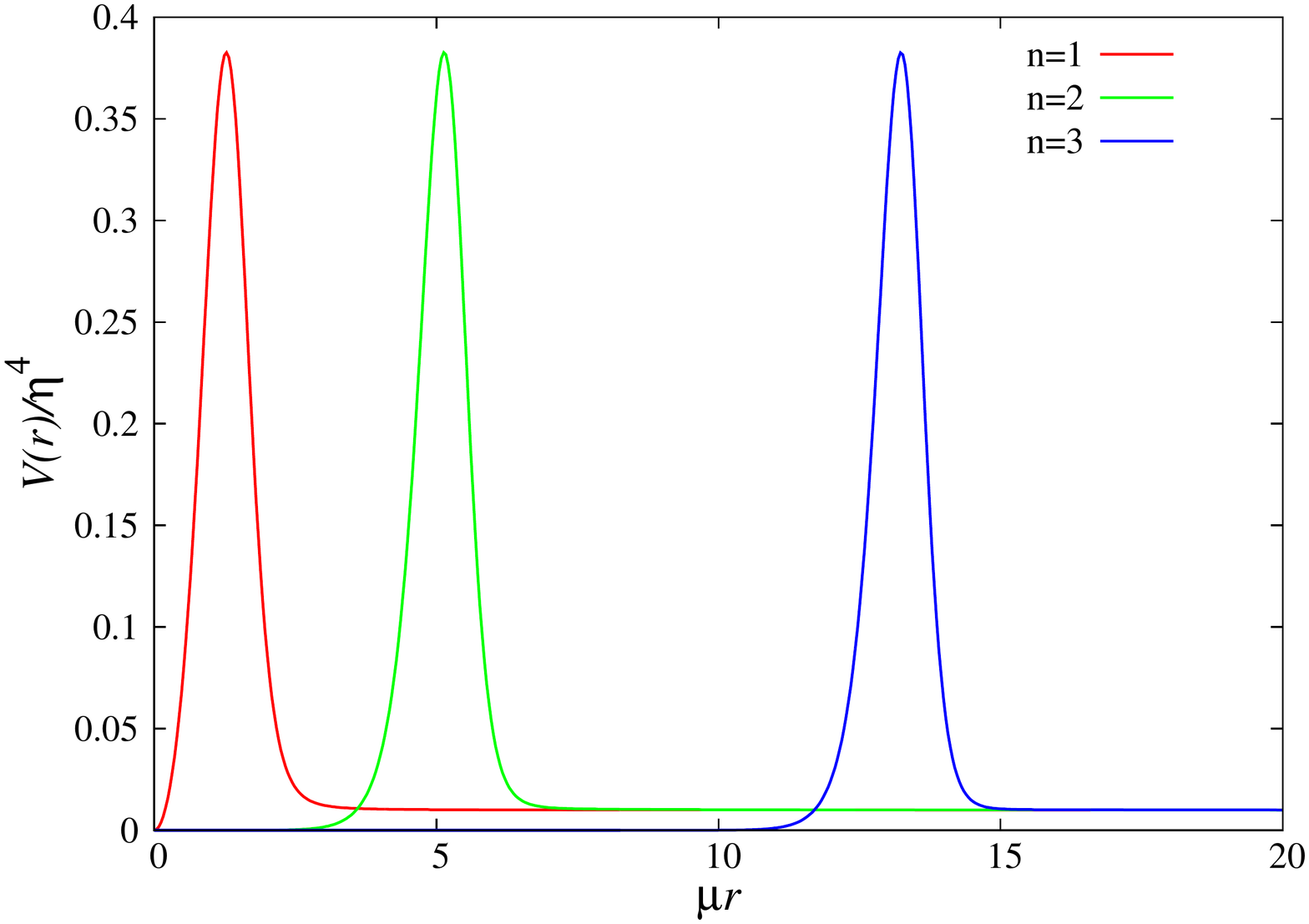}
}
\caption{The radial field configuration $R(r)$ ({\rm left}) and their
 potential energies ({\rm right}) with $\epsilon=0.1$,  $\zeta=4.0$, and $\delta=1.0$.
}
\label{fig:axialsolutions}
\end{figure}

The static axial-symmetric solutions can be obtained by solving
Eq.~(\ref{eq:R}) with the successive over-relaxation method
with the relaxation factor $\omega=1.0, 0.5$ and $0.3$ for $n=1,2,3$,
respectively. See the Appendix for details of our
numerical schemes. Using the boundary conditions
$R(r_{\rm b}) = R_{\rm min}$ and $R(0)=0$, we find stable solutions. In 
Fig.~\ref{fig:axialsolutions}, we plot the field configurations (left
panel) and the potential energies (right panel) of
the numerical solutions with $\epsilon=0.1, \zeta=4.0$ and $\delta=1.0$
for $n=1,2,3$. We confirmed that the field configurations are insensitive to the
position of boundary, $r_{\rm b}$, as long as it is sufficiently far from the
domain wall. At the region where the potential energy becomes a peak,
there is a domain wall which is the surface of the tube/cylinder. It is
found that the radius of the tube/cylinder depends on the winding
number, $n$, and the higher-winding solution tends to be
thicker. Moreover, the thickness of the domain wall is quite insensitive
to the winding number. 

%%%%%%%%%%%%%%%%%%%%%%%%%%%%%%%%%%%%%%%%%%%%%%%%%%
%%%%%%%%%%%%%%%%%%%%%%%%%%%%%%%%%%%%%%%%%%%%%%%%%%

%%%%%%%%%%%%%%%%%%%%%%%%%%%%%%%%%%%%%%%%%%%%%%%%%%%%%%%%%%%%%%%%%%%%%%%%%%%%%%%
%%%%%%%%%%%%%%%%%%%%%%%%%%%%%%%%%%%%%%%%%%%%%%%%%%%%%%%%%%%%%%%%%%%%%%%%%%%%%%%
\section{Analytic estimations}
\label{sec:ana}
%%%%%%%%%%%%%%%%%%%%%%%%%%%%%%%%%%%%%%%%%%%%%%%%%%%%%%%%%%%%%%%%%%%%%%%%%%%%%%%
%%%%%%%%%%%%%%%%%%%%%%%%%%%%%%%%%%%%%%%%%%%%%%%%%%%%%%%%%%%%%%%%%%%%%%%%%%%%%%%

%===============================
\subsection{A schematic illustration of stability of the tube-like string and bubble}
\label{subsec:thin}
%===============================

In the previous section, we demonstrate numerical solutions for the 
static tube-like string. Before we consider the detailed estimation of
their structure, we here give a rough approximation and clarify the
stability of those solutions by using a simple thin-wall
approximation. It is obvious that the stability of the tube-like string
depends on the difference in the energy density of the true and false
vacua 
%%%%%%%%%%%%%%%%%%%%%%%%%%%%%%%%%%%%%%%%%%%%%%%%%%%%%%%%%%%%%%%%%%%%
%
\begin{eqnarray}
{\tilde \epsilon} = V_{\rm false} - V_{\rm true} =V_{\rm min} \simeq \delta \mu^2\eta^2.
\end{eqnarray}
%
%%%%%%%%%%%%%%%%%%%%%%%%%%%%%%%%%%%%%%%%%%%%%%%%%%%%%%%%%%%%%%%%%%%%
When ${\tilde \epsilon} = 0$ $(\delta = 0)$, the vacua inside and
outside the string have the same energy. Then the tube-like string is
stable because it is supported by the topological reason. On the other
hand, when ${\tilde \epsilon}$ is large enough, the core energy density
wins out and the string grows and expands. Thus, one expects that there
exists a metastable tube-like string in an intermediate range for
${\tilde \epsilon}$. Let us clarify this by using the thin-wall
approximation where the wall of the string is thin compared to the core
and the exterior. The thin wall's contribution to the energy density can
be approximated  by the constant surface energy, $\sigma$. Then, the
tension of the string with radius $w$ is estimated by 
%%%%%%%%%%%%%%%%%%%%%%%%%%%%%%%%%%%%%%%%%%%%%%%%%%%%%%%%%%%%%%%%%%%%
%
\begin{eqnarray}
E(w) = n^2 C \log \frac{r_c}{w} + \sigma w  - {\tilde \epsilon} w^2,
\label{eq:schematic}
\end{eqnarray}
%
%%%%%%%%%%%%%%%%%%%%%%%%%%%%%%%%%%%%%%%%%%%%%%%%%%%%%%%%%%%%%%%%%%%%
where the first term is the contribution of the flux of the $U(1)$
global current with $r_c$ being a cutoff scale and 
$C\simeq 2 \pi \eta^2$ being a constant of a mass dimension 2, the
second term is of the wall and the last term is of the core of the
tube-like string. Here we assume ${\tilde \epsilon}, \sigma$, and $C$
are independent of $w$.
%%%%%%%%%%%%%%%%%%%%%%%%%%%%%%%%%%%%%%%%%%%%%%%%%%
\begin{figure}[!ht]
\centering{
  \includegraphics[width=7cm,bb=0 0 252 165]{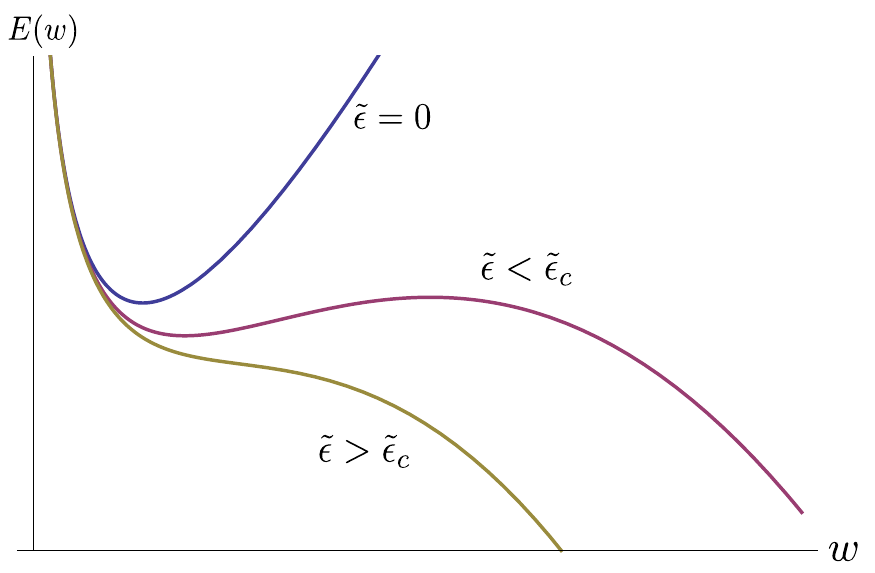}\qquad\qquad
  \includegraphics[width=7cm,bb=0 0 247 161]{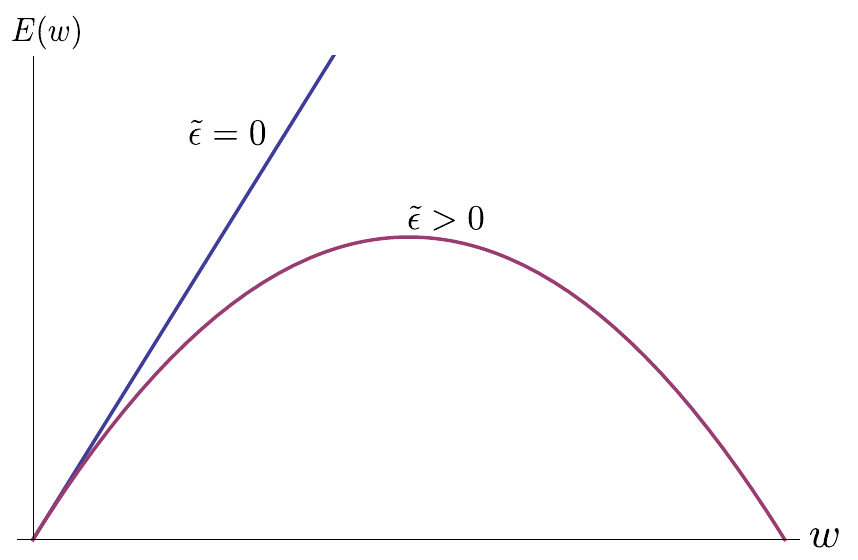}
}
\caption{A string energy (left) and a bubble energy (right) as functions of the core radius $w$.}
\label{fig:schematic}
\end{figure}
%%%%%%%%%%%%%%%%%%%%%%%%%%%%%%%%%%%%%%%%%%%%%%%%%%
The dependence of the string energy on ${\tilde \epsilon}$ is shown in
Fig.~\ref{fig:schematic}. When ${\tilde \epsilon }= 0$, as we explained,
there exists a local minimum, which implies  that there exists a stable
tube-like string. In the region 
$0 < {\tilde \epsilon} < {\tilde \epsilon}_c$
with ${\tilde \epsilon}_c = \frac{\sigma^2}{8Cn^2}$, there exist a local
minimum and a global maximum at
%%%%%%%%%%%%%%%%%%%%%%%%%%%%%%%%%%%%%%%%%%%%%%%%%%%%%%%%%%%%%%%%%%%%
%
\begin{equation}
w_{\rm min} = \frac{\sigma -\sqrt{\sigma ^2-8 C n^2 {\tilde\epsilon }}
}{4 {\tilde \epsilon }},\quad
w_{\rm max} = \frac{\sigma +\sqrt{\sigma ^2-8 C n^2 {\tilde \epsilon}
  } }{4 {\tilde \epsilon }}.
\end{equation}
%
%%%%%%%%%%%%%%%%%%%%%%%%%%%%%%%%%%%%%%%%%%%%%%%%%%%%%%%%%%%%%%%%%%%%
In this case the tube-like strings exist but they are metastable. The
tube-like string whose core radius is larger than $w_{\rm max}$ is
unstable. When ${\tilde \epsilon}$ exceeds the critical value, 
${\tilde \epsilon} > {\tilde \epsilon}_c$, there are no local minima at
all. Namely, the string is unstable and the string core expands
infinitely. In this paper the metastable strings are concerned. Although
they are metastable as static configurations, they will become unstable
in some dynamical processes such as string collision, annihilation and
reconnections. 

From Eq.~(\ref{eq:schematic}) with $n$ being zero, the fate of the
two-dimensional bubble can be also clarified. As shown in
Fig.~\ref{fig:schematic}, the bubble never grows if 
${\tilde \epsilon }=0$. Once the positive ${\tilde \epsilon }$ is turned
on, the bubble becomes metastable. The critical radius of the bubble is 
$w_c=\frac{\sigma}{2{\tilde \epsilon}}$.
The bubble lager than $w_c$ expands infinitely and the true vacua inside
the core wins out. There are no metastable tube-like solutions. 

The thin-wall approximation explained above is simple and gives an
insight about the stability problem. Nevertheless, it is of only limited
accuracy and we need a better analysis to get more quantitative results
beyond the qualitative properties. To go beyond the thin-wall
approximation, we will work out another analytical study in the next
subsections.

%===============================
\subsection{Thickness of walls}
%===============================

Now let us go beyond the thin-wall approximation and consider the
stability of the tube-like solution in more detail. From the numerical
calculations of stable static solutions, we found that the tube-like
solution is well characterized by the radius of cylinders, $w$, the winding number, 
and the thickness of the walls, $d$. Since the thickness of the wall is
quite insensitive to other two parameters, it is possible to estimate
$d$ from the variation of the total energy $E$ with respect to $d$, and
it can be done by considering only the case of $n=1$.

We here simply approximate the solution with $n=1$ as the following
piecewise linear function
%%%%%%%%%%%%%%%%%%%%%%%%%%%%%%%%%%%%%%%%%%%%%%%%%%%%%%%%%%%%%%%%%%%%
%
\begin{equation}
 R(r) \approx 
\begin{cases}
  \displaystyle \frac{r}{d}R_{\rm min} & r < d \\
  R_{\rm min} & d < r
\end{cases},
\end{equation}
%
%%%%%%%%%%%%%%%%%%%%%%%%%%%%%%%%%%%%%%%%%%%%%%%%%%%%%%%%%%%%%%%%%%%%
where we set $w=0$, see Fig.~\ref{fig:axialsolutions}. Then, we
approximate the volume integral of the potential energy and gradient
energies as 
%%%%%%%%%%%%%%%%%%%%%%%%%%%%%%%%%%%%%%%%%%%%%%%%%%%%%%%%%%%%%%%%%%%%
%
\begin{align}
 E_{\rm potential} &= \pi d^2(V_{\rm max}-V_{\rm min}), \\
 E_{\rm grad,r} &= \pi\eta^2R_{\rm min}^2,\\
 E_{\rm grad,\theta} &= \pi \eta^2R_{\rm min}^2\left(1+2\log\frac{r_c}{d}\right).
\end{align}
%
%%%%%%%%%%%%%%%%%%%%%%%%%%%%%%%%%%%%%%%%%%%%%%%%%%%%%%%%%%%%%%%%%%%%
The derivative of the summation of these energy
components with respect to $d$ gives the desired value of $d$,
%%%%%%%%%%%%%%%%%%%%%%%%%%%%%%%%%%%%%%%%%%%%%%%%%%%%%%%%%%%%%%%%%%%%
%
\begin{equation}
 \frac{\partial E}{\partial d} = 2\pi d(V_{\rm max}-V_{\rm min})
    - 2\pi \eta^2\frac{R_{\rm min}^2}{d}=0
 \quad \Longrightarrow \quad
   d = \frac{\eta R_{\rm min}}{\sqrt{V_{\rm max}-V_{\rm min}}}
     = \frac{3}{2\eta\epsilon\zeta^2}\sqrt{\frac{c+2}{c^3}}.
 \label{eq:d}
\end{equation}
%
%%%%%%%%%%%%%%%%%%%%%%%%%%%%%%%%%%%%%%%%%%%%%%%%%%%%%%%%%%%%%%%%%%%%

%===============================
\subsection{Upper bound of winding number}
%===============================

We consider how the radius of the cylinder is determined and the upper
limit of the winding number with which the static solution exists. 
As shown in Fig.~\ref{fig:axialsolutions}, for $n \ge 2$, there are two
parameters characterizing a cylinder, the radius $w$ and the thickness
$d$ of the surface. However, as shown in the previous subsections, the
latter one is not sensitive to the winding number and is 
determined from $V_{\rm max} - V_{\rm min}$.
Hence we here focus on the radius $w$ of the cylinder while the width of
the wall $d$ is assumed to be the same as one given in
Eq.~(\ref{eq:d}). 

\begin{figure}[!ht]
\centering{
  \includegraphics[width=8cm,bb=0 0 792 612]{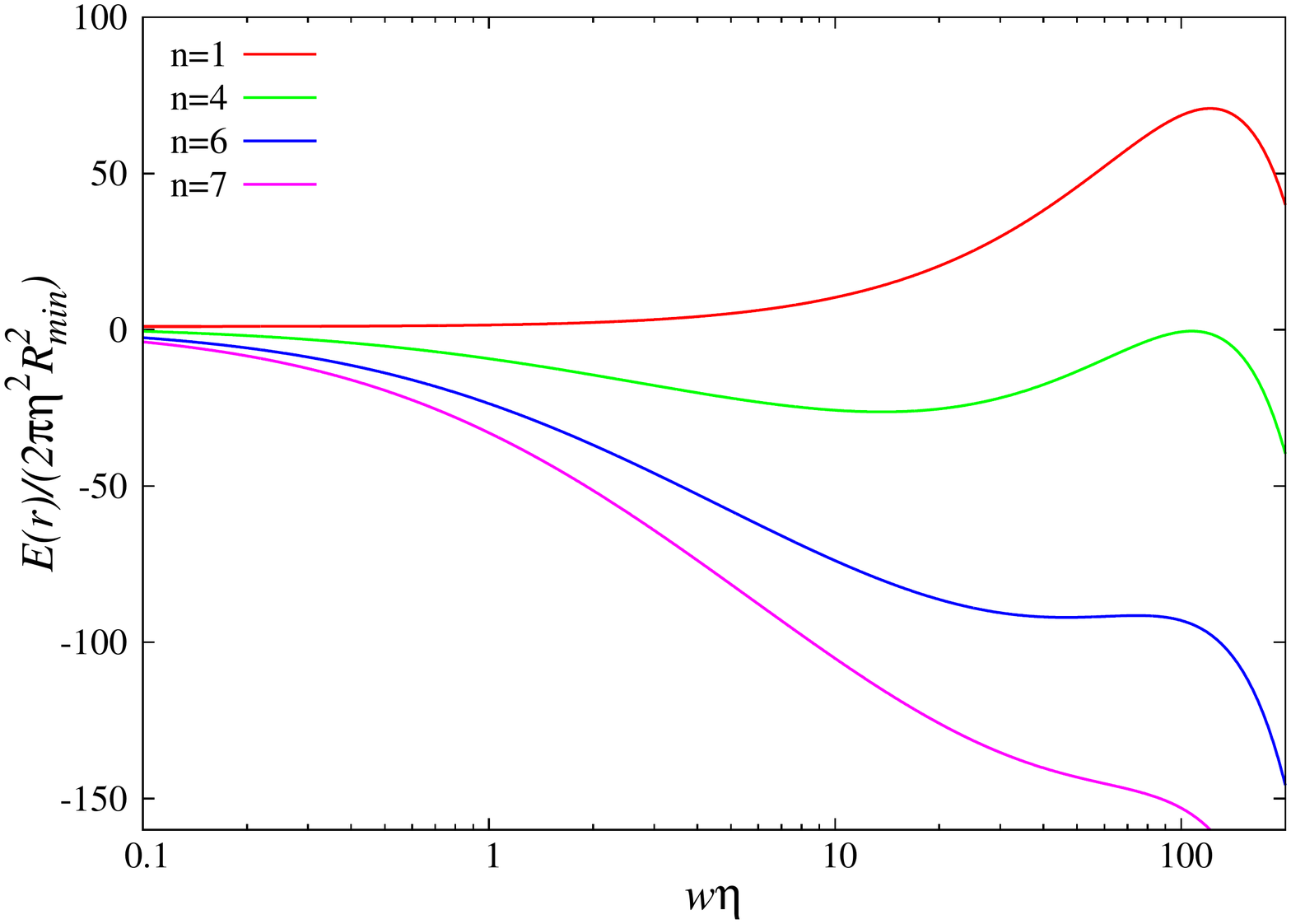}
  \includegraphics[width=8cm,bb=0 0 792 612]{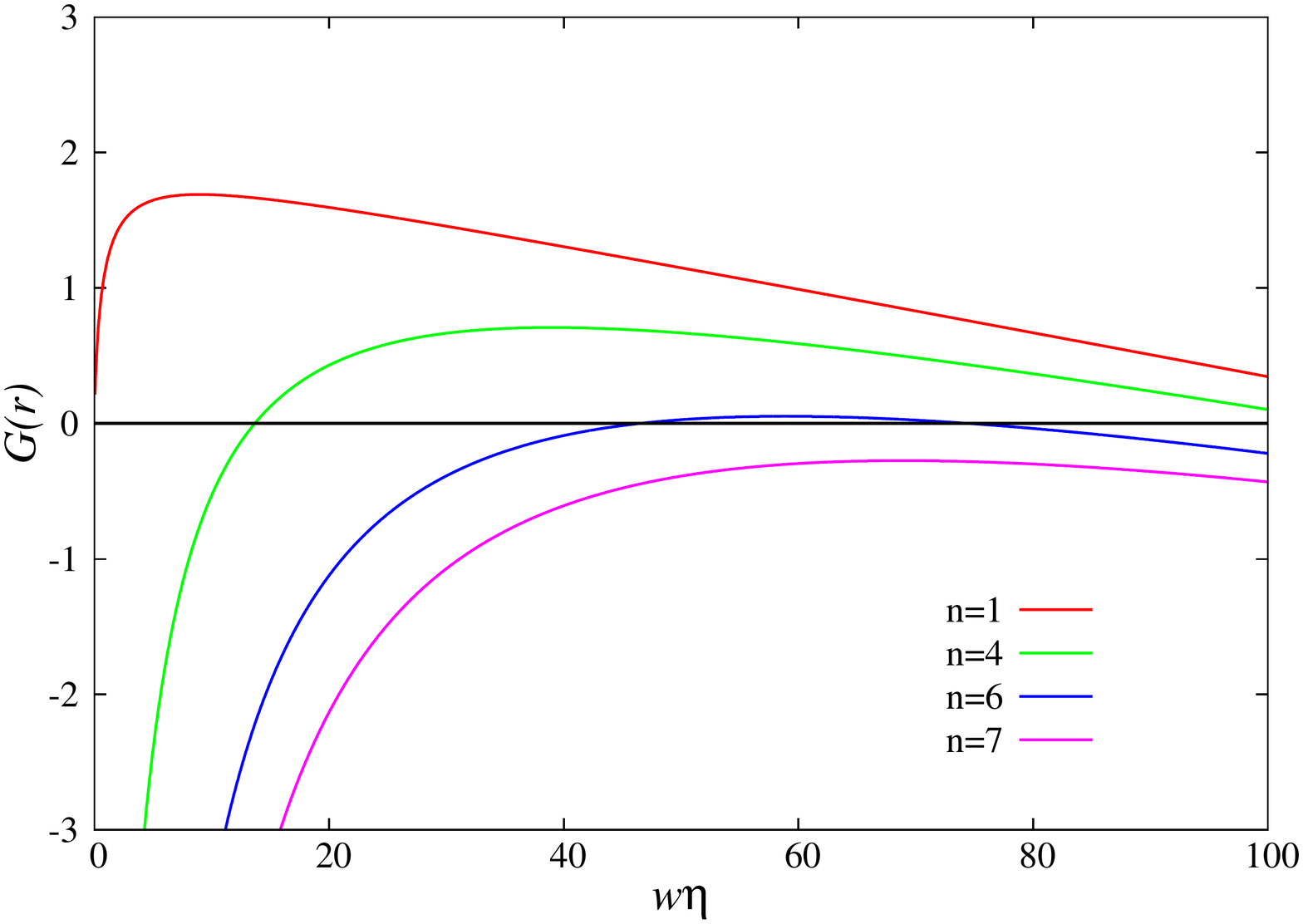}
}
\caption{An example of $E(w)/(2\pi\eta^2 R_{\rm min}^2)$ ({\it left}) and $G(w)$
 ({\it right}) with specific values of $\epsilon=0.1, \zeta=4.0,$
 and $\delta=1.0$.}
\label{fig:example1}
\end{figure}

We simply approximate the solution and the potential as
%%%%%%%%%%%%%%%%%%%%%%%%%%%%%%%%%%%%%%%%%%%%%%%%%%%%%%%%%%%%%%%%%%%%
%
\begin{equation}
 R(r) \approx 
\begin{cases}
  0 & \\
  \displaystyle \frac{r-w}{d}R_{\rm min} & \\
  R_{\rm min} & 
\end{cases},
 V(r) \approx 
\begin{cases}
  0 & r<w \\
  V_{\rm max} & w \leq r < w+d \\
  V_{\rm min} & w+d < r
\end{cases}.
\end{equation}
%
%%%%%%%%%%%%%%%%%%%%%%%%%%%%%%%%%%%%%%%%%%%%%%%%%%%%%%%%%%%%%%%%%%%%
The volume integral of the potential energy is approximated as
%%%%%%%%%%%%%%%%%%%%%%%%%%%%%%%%%%%%%%%%%%%%%%%%%%%%%%%%%%%%%%%%%%%%
%
\begin{equation}
 E_{\rm potential} = -V_{\rm min}\pi w^2 + \pi d(2w+d)(V_{\rm max}-V_{\rm min}),
 \label{eq:Ep}
\end{equation}
%
%%%%%%%%%%%%%%%%%%%%%%%%%%%%%%%%%%%%%%%%%%%%%%%%%%%%%%%%%%%%%%%%%%%%
where we considered only the deviation from $V_{\rm min}$.
On the other hand, that of the gradient energy can be divided into the
radial and angular parts,
%%%%%%%%%%%%%%%%%%%%%%%%%%%%%%%%%%%%%%%%%%%%%%%%%%%%%%%%%%%%%%%%%%%%
%
\begin{align}
 E_{\rm grad,r} &= 2\pi\eta^2\int_0^\infty\! R'(r){}^2 \,rdr
     = \pi\eta^2R_{\rm min}^2\left(2\frac{w}{d}+1\right),\\
 E_{\rm grad,\theta} &= 2\pi n^2\eta^2\int_0^\infty\! \frac{R^2}{r^2} \,rdr
     = 2\pi n^2\eta^2R_{\rm min}^2\left[
          \frac{1}{2}-\frac{w}{d}+\frac{w^2}{d^2}\log\left(1+\frac{d}{w}\right)
          + \log\left(\frac{r_c}{w+d}\right)
      \right],
\end{align}
%
%%%%%%%%%%%%%%%%%%%%%%%%%%%%%%%%%%%%%%%%%%%%%%%%%%%%%%%%%%%%%%%%%%%%
where $r_c$ is the cut-off length, $r_c\to\infty$, and 
it should be properly regularized.
Then the total energy becomes
%%%%%%%%%%%%%%%%%%%%%%%%%%%%%%%%%%%%%%%%%%%%%%%%%%%%%%%%%%%%%%%%%%%%
%
\begin{equation}
 E = E_{\rm potential} + E_{{\rm grad},r} + E_{{\rm grad},\theta}.
\end{equation}
%
%%%%%%%%%%%%%%%%%%%%%%%%%%%%%%%%%%%%%%%%%%%%%%%%%%%%%%%%%%%%%%%%%%%%
Let us look for the values of $w$ to minimize the total energy.
One can easily see that there is no global
minimum of $E$ since the term proportional to $-w^2$ in Eq.~(\ref{eq:Ep})
implies that $E\to -\infty$ for $w\to \infty$.
Instead, we investigate whether there is a local minimum in the region, $w>0$.
The derivative of $E$ with respect to $w$ is calculated as
%%%%%%%%%%%%%%%%%%%%%%%%%%%%%%%%%%%%%%%%%%%%%%%%%%%%%%%%%%%%%%%%%%%%
%
\begin{equation}
 \begin{aligned}
 \frac{dE}{dw} &= \frac{2\pi\eta^2R_{\rm min}^2}{d}G(w), \\
 G(w) &\equiv -2p\frac{w}{d}+2 + 2n^2\left[-1 + \frac{w}{d}\log\left(1+\frac{d}{w}\right)\right],
 \end{aligned}
  \label{eq:G}
\end{equation}
%
%%%%%%%%%%%%%%%%%%%%%%%%%%%%%%%%%%%%%%%%%%%%%%%%%%%%%%%%%%%%%%%%%%%%
where
%%%%%%%%%%%%%%%%%%%%%%%%%%%%%%%%%%%%%%%%%%%%%%%%%%%%%%%%%%%%%%%%%%%%
%
\begin{equation}
 p \equiv \frac{d^2V_{\rm min}}{2\eta^2R_{\rm min}^2}
     = \frac{V_{\rm  min}}{2(V_{\rm max}-V_{\rm min})}.
 \label{eq:def_p}
\end{equation}
%
%%%%%%%%%%%%%%%%%%%%%%%%%%%%%%%%%%%%%%%%%%%%%%%%%%%%%%%%%%%%%%%%%%%%
What we have to do is to find $w$ to satisfy $G(w)=0$ for $w>0$. 
We plot $E(w)$ and $G(w)$ with specific values of 
$\epsilon, \zeta$ and $\delta$ in Fig.~\ref{fig:example1}. 
From the right panel of this figure, it is found that there are two
zero-points for $n\leq 6$, and the smaller one is the desired value of
$w$, at which $E$ is locally minimized, and the other zero-point gives
the local maximum of $E$ (see the corresponding lines in the left
panel). Note that the local minimum of $n=1$ is $w=0$, which is
consistent with the numerical solution in Fig.~\ref{fig:axialsolutions}.

In order for these zero-points to exist, the local maximum of $G(w)$
should be positive. Actually, $G(w)$ for $n=7$ is always negative, and
in the left panel of Fig.~\ref{fig:example1} the line for $E(w)$ with
$n=7$ has no local minimum, which indicates the cylinder solution with
$n=7$ is unstable. To proceed this calculation, we assume $d/w\ll 1$ to
expand the logarithm function. Then we find that $G(w)$ has a local maximum,
%%%%%%%%%%%%%%%%%%%%%%%%%%%%%%%%%%%%%%%%%%%%%%%%%%%%%%%%%%%%%%%%%%%%
%
\begin{equation}
G(w_c) \simeq  2-2n \sqrt{2p},\quad
w_c \simeq \frac{nd}{\sqrt{2p}}.
\end{equation}
%
%%%%%%%%%%%%%%%%%%%%%%%%%%%%%%%%%%%%%%%%%%%%%%%%%%%%%%%%%%%%%%%%%%%%
Therefore the condition on the winding number required for the local minimum of $E$ is
%%%%%%%%%%%%%%%%%%%%%%%%%%%%%%%%%%%%%%%%%%%%%%%%%%%%%%%%%%%%%%%%%%%%
%
\begin{equation}
G(w_c) > 0 \quad \Longrightarrow \quad
n < \frac{1}{\sqrt{2p}} = \sqrt{\frac{V_{\rm max}-V_{\rm min}}{V_{\rm min}}}.
\label{eq:winding_limit}
\end{equation}
%
%%%%%%%%%%%%%%%%%%%%%%%%%%%%%%%%%%%%%%%%%%%%%%%%%%%%%%%%%%%%%%%%%%%%

%%%%%%%%%%%%%%%%%%%%%%%%%%%%%%%%%%%%%%%%%%%%%%%%%%
%%%%%%%%%%%%%%%%%%%%%%%%%%%%%%%%%%%%%%%%%%%%%%%%%%

%===============================
\subsection{Instability of spherical bubble and critical volume}
%===============================

In subsection \ref{subsec:thin} we have seen that the two-dimensional bubble is
unstable and it shrinks (expands) when its radius is smaller (larger)
than the critical value. Here, we study the spherically symmetric
configuration (three-dimensional bubble) of the scalar field separated
by the domain wall. It is also expected to be always unstable.
After two cylinders collide with each other, there appears a spherical
object at the impact point. To study the instability of such an object,
we consider the following ansatz in the spherical coordinate,
%%%%%%%%%%%%%%%%%%%%%%%%%%%%%%%%%%%%%%%%%%%%%%%%%%%%%%%%%%%%%%%%%%%%
%
\begin{equation}
 R(r) \approx 
\begin{cases}
  0 & \\
  \displaystyle \frac{r}{d}R_{\rm min} & \\
  R_{\rm min} & 
\end{cases}, 
 V(r) \approx 
\begin{cases}
  0 & r < w\\
  V_{\rm max} & w < r < w+d \\
  V_{\rm min} & d < r
\end{cases},
\end{equation}
%
%%%%%%%%%%%%%%%%%%%%%%%%%%%%%%%%%%%%%%%%%%%%%%%%%%%%%%%%%%%%%%%%%%%%
with assuming that $d$ is given by Eq.~\eqref{eq:d}. 
We simply assume that the field is homogeneous in the azimuthal and the
polar directions\footnote{Strictly speaking, the separation of variables
is not justified, since it is impossible to expand $X$ in the
spherical harmonics because of the nonlinear terms in the potential.}.
The potential and radial gradient energy are approximated as
%%%%%%%%%%%%%%%%%%%%%%%%%%%%%%%%%%%%%%%%%%%%%%%%%%%%%%%%%%%%%%%%%%%%
%
\begin{align}
 E_{\rm potential} &= -\frac{4}{3}\pi w^3V_{\rm min}
      + \frac{4}{3}\pi\{(w+d)^3-w^3\}(V_{\rm max}-V_{\rm min}), \\
 E_{\rm grad,r} &= \frac{4}{3}\pi\left(\frac{R_{\rm min}}{d}\right)^2\{(w+d)^3-w^3\}\eta^2,
\end{align}
%
%%%%%%%%%%%%%%%%%%%%%%%%%%%%%%%%%%%%%%%%%%%%%%%%%%%%%%%%%%%%%%%%%%%%
and thus the total energy becomes
%%%%%%%%%%%%%%%%%%%%%%%%%%%%%%%%%%%%%%%%%%%%%%%%%%%%%%%%%%%%%%%%%%%%
%
\begin{equation}
 E(w) = \frac{4}{3}\pi\left[
    -w^3V_{\rm min} + 2\{(w+d)^3-w^3\}  (V_{\rm max} - V_{\rm min})
  \right],
\end{equation}
%
%%%%%%%%%%%%%%%%%%%%%%%%%%%%%%%%%%%%%%%%%%%%%%%%%%%%%%%%%%%%%%%%%%%%
where we used Eq.~(\ref{eq:d}) to eliminate $R_{\rm min}$.
The function $E(w)$ has the only peak at $w=w_c$ in $w>0$, and 
the critical radius $w_c$ is given by solving $dE/dw=0$,
%%%%%%%%%%%%%%%%%%%%%%%%%%%%%%%%%%%%%%%%%%%%%%%%%%%%%%%%%%%%%%%%%%%%
%
\begin{equation}
 w_c = \frac{1+\sqrt{1+p}}{p}d, \label{eq:critical_radius}
\end{equation}
%
%%%%%%%%%%%%%%%%%%%%%%%%%%%%%%%%%%%%%%%%%%%%%%%%%%%%%%%%%%%%%%%%%%%%
where $p$ is defined in Eq.~(\ref{eq:def_p}). For $w<w_c$ the radius of
the sphere starts to shrink since $dE/dw>0$. On the other hand, for
$w>w_c$ the radius goes to the positive infinity since $dE/dw<0$.
That is, if the volume of the spherical object is larger than 
the critical volume $4\pi w_c^3/3$, this grows infinitely.

%%%%%%%%%%%%%%%%%%%%%%%%%%%%%%%%%%%%%%%%%%%%%%%%%%
\begin{figure}[!ht]
\centering{
  \includegraphics[width=8cm, bb=0 0 350 350]{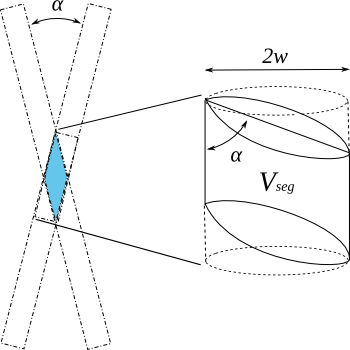}
}
\caption{The volume colliding with another cylinder.}
\label{fig:segment}
\end{figure}
%%%%%%%%%%%%%%%%%%%%%%%%%%%%%%%%%%%%%%%%%%%%%%%%%%

Now we give an analytical estimate of the (in)stability of the colliding cylinders. 
Let us consider the situation that two cylinders collide with the
collision angle $\alpha$ as shown in Fig.~\ref{fig:segment}. Assuming that 
a spherical object is produced via the fusion of two cylinders after collision and 
its volume is equal to
the total volume of the two colliding segments of cylinders, $2V_{\rm seg}$,
the initial volume of the spherical object is calculated as
%%%%%%%%%%%%%%%%%%%%%%%%%%%%%%%%%%%%%%%%%%%%%%%%%%%%%%%%%%%%%%%%%%%%
%
\begin{equation}
 V_{\rm total} = 2V_{\rm seg} = \frac{4\pi w^3}{\sin \alpha}.
\end{equation}
%
%%%%%%%%%%%%%%%%%%%%%%%%%%%%%%%%%%%%%%%%%%%%%%%%%%%%%%%%%%%%%%%%%%%%
Then the condition for the sphere to grow infinitely is given by
%%%%%%%%%%%%%%%%%%%%%%%%%%%%%%%%%%%%%%%%%%%%%%%%%%%%%%%%%%%%%%%%%%%%
%
\begin{equation}
 \frac{4\pi}{3}w_c^3 < V_{\rm total} \;\;\Longrightarrow\;\;
  \sin\alpha < 3\left(\frac{w}{w_c}\right)^3,
\end{equation}
%
%%%%%%%%%%%%%%%%%%%%%%%%%%%%%%%%%%%%%%%%%%%%%%%%%%%%%%%%%%%%%%%%%%%%
where $w$ is obtained by solving $G(w)=0$ in Eq.~(\ref{eq:G}), and hence
assuming $d/w \ll 1$, this is calculated as
%%%%%%%%%%%%%%%%%%%%%%%%%%%%%%%%%%%%%%%%%%%%%%%%%%%%%%%%%%%%%%%%%%%%
%
\begin{equation}
 w = \frac{1-\sqrt{1-2n^2p}}{2p}d.
\end{equation}
%
%%%%%%%%%%%%%%%%%%%%%%%%%%%%%%%%%%%%%%%%%%%%%%%%%%%%%%%%%%%%%%%%%%%%
As a result, we obtain the upper limit of the collision angle so that
the spherical object created at the impact point can grow infinitely,
%%%%%%%%%%%%%%%%%%%%%%%%%%%%%%%%%%%%%%%%%%%%%%%%%%%%%%%%%%%%%%%%%%%%
%
\begin{equation}
 \sin \alpha < \frac{3}{8}\left(\frac{1-\sqrt{1-2n^2p}}{1+\sqrt{1+p}}\right)^3.
\end{equation}
%
%%%%%%%%%%%%%%%%%%%%%%%%%%%%%%%%%%%%%%%%%%%%%%%%%%%%%%%%%%%%%%%%%%%%
For example, $\epsilon=0.1, \zeta=4.0$ and $\delta=1.0$ give the upper
limit, $\alpha<7.78\times 10^{-6}$, for $n=2$, and 
$\alpha<2.50\times 10^{-2}$ for $n=6$.

%%%%%%%%%%%%%%%%%%%%%%%%%%%%%%%%%%%%%%%%%%%%%%%%%%%%%%%%%%%%%%%%%%%%%%%%%%%%%%%
%%%%%%%%%%%%%%%%%%%%%%%%%%%%%%%%%%%%%%%%%%%%%%%%%%%%%%%%%%%%%%%%%%%%%%%%%%%%%%%
\section{Simulation setup}
\label{sec:setup}
%%%%%%%%%%%%%%%%%%%%%%%%%%%%%%%%%%%%%%%%%%%%%%%%%%%%%%%%%%%%%%%%%%%%%%%%%%%%%%%
%%%%%%%%%%%%%%%%%%%%%%%%%%%%%%%%%%%%%%%%%%%%%%%%%%%%%%%%%%%%%%%%%%%%%%%%%%%%%%%

We explore the (in)stability of collision processes of two 
meta-stable strings. The individual string is given by the axial-symmetric
solution which is numerically obtained by solving Eq.~(\ref{eq:fieldeq})
without the time-derivative term, corresponding to Eq.~(\ref{eq:R}).
This procedure is nothing but the one-dimensional boundary value
problem, and is carried out with the Gauss-Seidel method.
Each of the resultant strings is put on the x-z surface parallel to each
other separated a bit. The strings are Lorentz-boosted to collide with
each other at an angle with $\alpha$ measured on the x-z surface
including the origin (to be the impact point when colliding). 
The free model parameters are $\epsilon$ and $\zeta$, which control the
strength of the self-coupling  and the energy scale of the phase
transition, respectively. The fiducial parameters are listed in
Table~\ref{table:param}.
Note that we choose $\zeta=2.2$ for numerical simulations, while
$\zeta=4.0$ has been used so far, to enhance the instability. 

The initial separation between the strings is fixed to be $20\eta^{-1}$, 
while the radius of the string is about $5\eta^{-1}$ for the fiducial
choice of parameters. According to Ref.~\cite{Vilenkin}, the
superposition of the two strings is given by 
%%%%%%%%%%%%%%%%%%%%%%%%%%%%%%%%%%%%%%%%%%%%%%%%%%%%%%%%%%%%%%%%%%%%
%
\begin{equation}
 X = \frac{X_1X_2}{\eta R_{\rm min}},
\end{equation}
%
%%%%%%%%%%%%%%%%%%%%%%%%%%%%%%%%%%%%%%%%%%%%%%%%%%%%%%%%%%%%%%%%%%%%
where the denominator is determined by the dimension analysis and the
fact that $|X|\to \eta R_{\rm min}$ far away from two strings.

Then we solve Eq.~(\ref{eq:fieldeq}) in the three-dimensional cartesian
coordinate with the Neumann conditions on the boundaries, 
$\partial X/\partial n^i|_{\rm boundary}=0$, where $n^i$ is the normal
vector to each boundary. We use the Leap-frog method for the time domain
and approximate the spatial derivatives by the 2nd-order central finite
difference. The simulations are stopped at $t=L/2$, when
the information at the impact point arrives at the nearest boundary of
the computational domain.

%%%%%%%%%%%%%%%%%%%%%%%%%%%%%%%%%%%%%%%%%%%%%%%%%%
\begin{table}[!ht]
\begin{tabular}{|c|c|}
\hline
\multicolumn{2}{|c|}{model parameters} \\
\hline
$\epsilon$ & 0.1 \\
$\zeta$ & 2.2 \\
$\delta$ & 1.0 \\
$n_1$ & 1 \\
$n_2$ & 1 \\
\hline 
\end{tabular}
\begin{tabular}{|c|c|c|}
\hline
\multicolumn{3}{|c|}{simulation parameters} \\
\hline 
grid size     & $N$ & $480^3$ \\
box size      & $L$ & 240$\eta^{-1}$ \\
time interval & $\Delta t$ & $0.2\eta^{-1}$ \\
total steps   & $N_{\rm s}$ & 600 \\
simulation time  & $t_f$ & $120\eta^{-1}$ \\
\hline
\end{tabular}
\caption{The fiducial parameters.}
\label{table:param}
\end{table}
%%%%%%%%%%%%%%%%%%%%%%%%%%%%%%%%%%%%%%%%%%%%%%%%%%

%%%%%%%%%%%%%%%%%%%%%%%%%%%%%%%%%%%%%%%%%%%%%%%%%%%%%%%%%%%%%%%%%%%%%%%%%%%%%%%
%%%%%%%%%%%%%%%%%%%%%%%%%%%%%%%%%%%%%%%%%%%%%%%%%%%%%%%%%%%%%%%%%%%%%%%%%%%%%%%
\section{Results}
\label{sec:results}
%%%%%%%%%%%%%%%%%%%%%%%%%%%%%%%%%%%%%%%%%%%%%%%%%%%%%%%%%%%%%%%%%%%%%%%%%%%%%%%
%%%%%%%%%%%%%%%%%%%%%%%%%%%%%%%%%%%%%%%%%%%%%%%%%%%%%%%%%%%%%%%%%%%%%%%%%%%%%%%

\begin{figure}[!ht]
\centering{
  \includegraphics[bb=0 0 290 290,width=3.3cm]{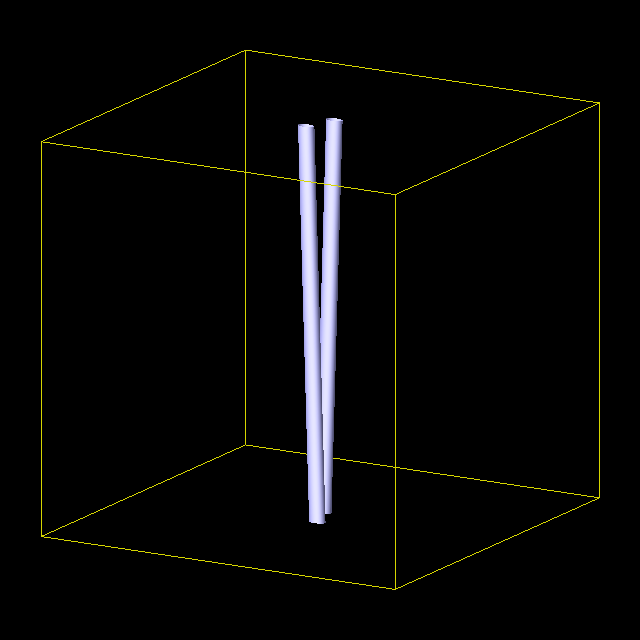}
  \includegraphics[bb=0 0 290 290,width=3.3cm]{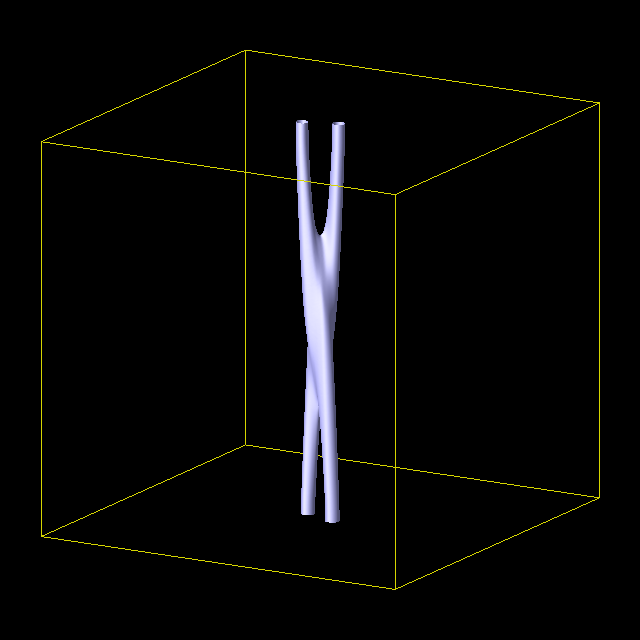}
  \includegraphics[bb=0 0 290 290,width=3.3cm]{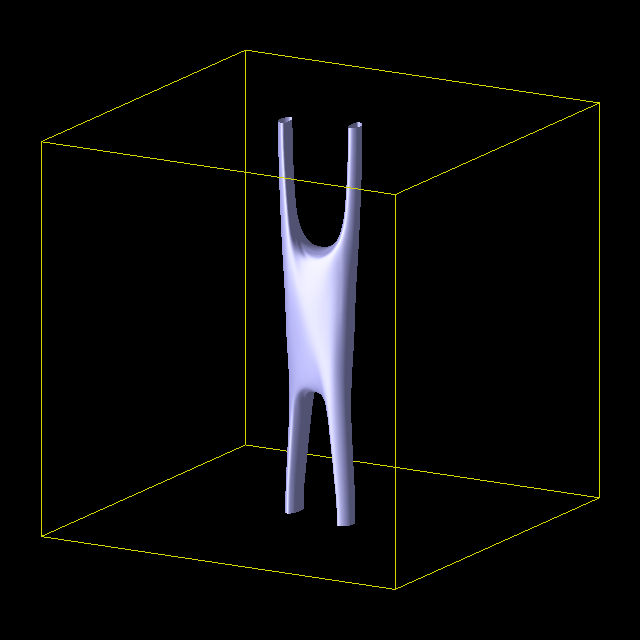}
  \includegraphics[bb=0 0 290 290,width=3.3cm]{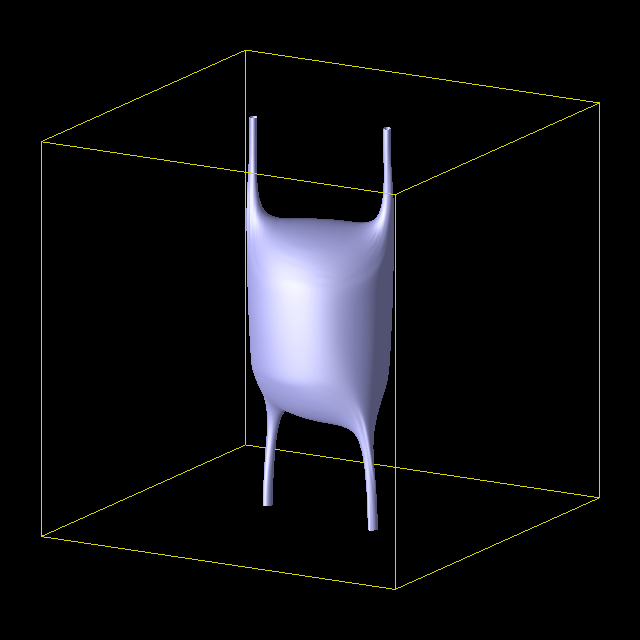}
  \includegraphics[bb=0 0 290 290,width=3.3cm]{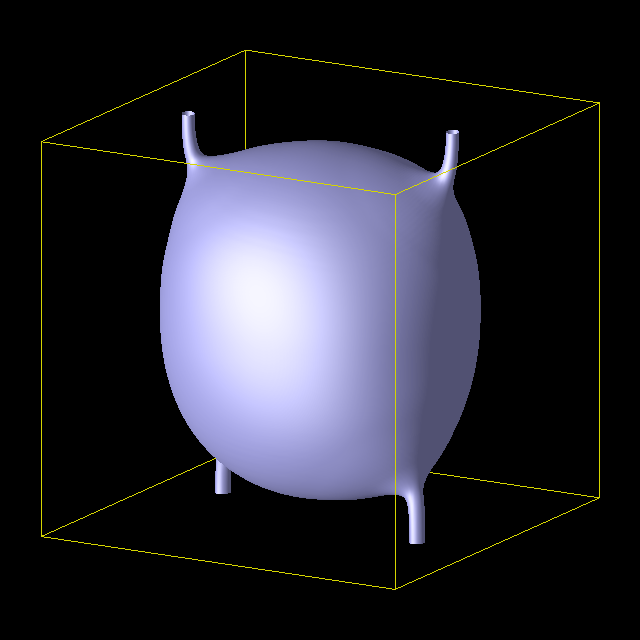}
}
\caption{[P1] Failed reconnection of parallel string pair with
 $(v,\alpha)=(0.82c,0.02\pi)$. See \cite{animation} for full
 animations.} 
\label{fig:3D_P1}
\end{figure}

\begin{figure}[!ht]
\centering{
  \includegraphics[width=3.3cm,bb=0 0 290 290]{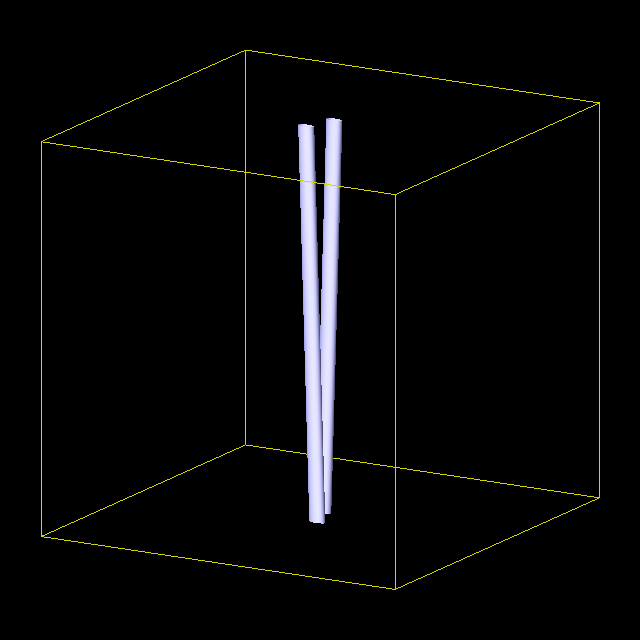}
  \includegraphics[width=3.3cm,bb=0 0 290 290]{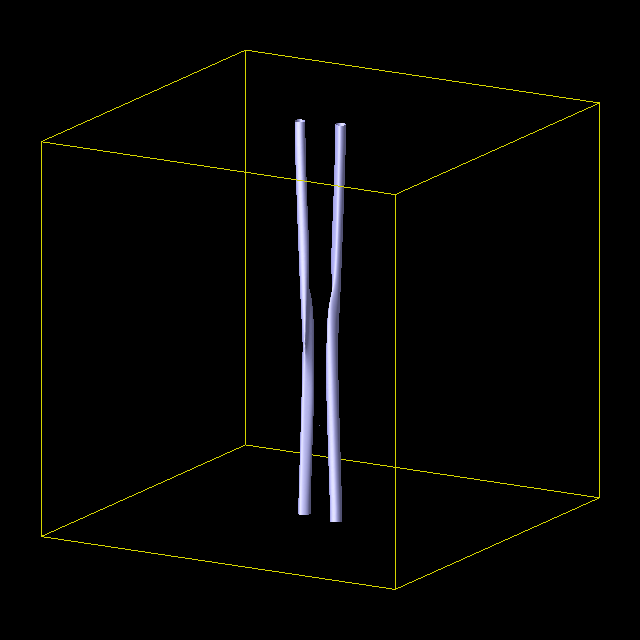}
  \includegraphics[width=3.3cm,bb=0 0 290 290]{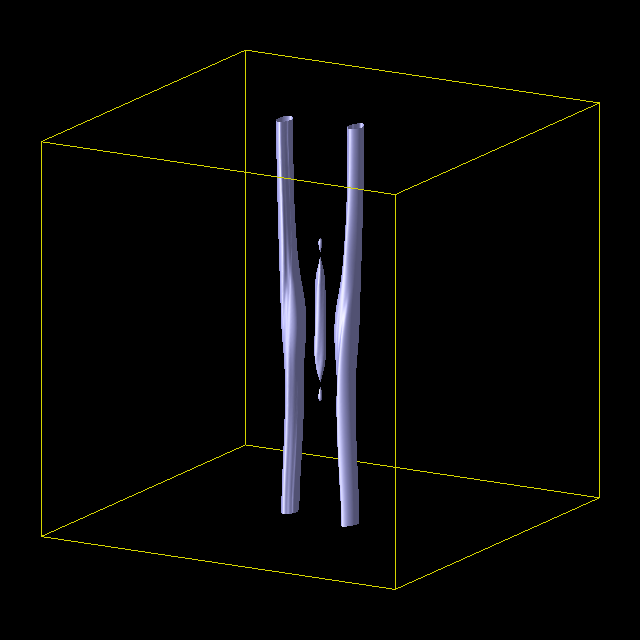}
  \includegraphics[width=3.3cm,bb=0 0 290 290]{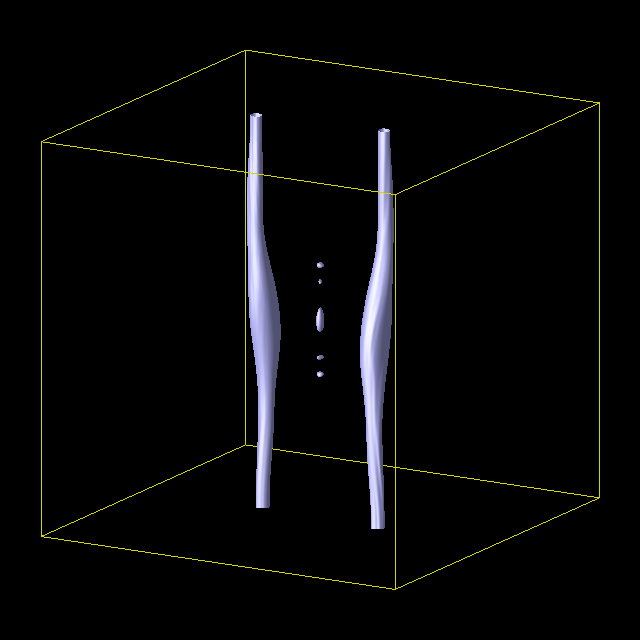}
  \includegraphics[width=3.3cm,bb=0 0 290 290]{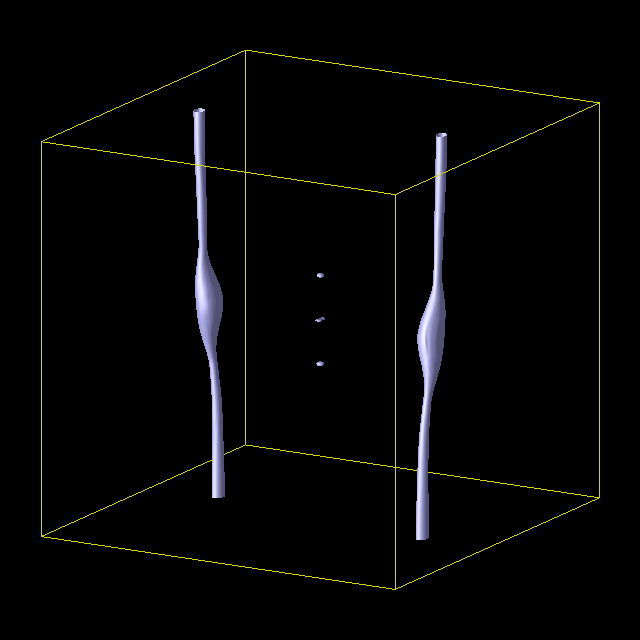}
}
\caption{[P2] Successful reconnection of parallel string pair with
 $(v,\alpha)=(0.90c,0.02\pi)$. See \cite{animation} for full
 animations.} 
\label{fig:3D_P2}
\end{figure}

\begin{figure}[!ht]
\centering{
  \includegraphics[width=3.3cm,bb=0 0 290 290]{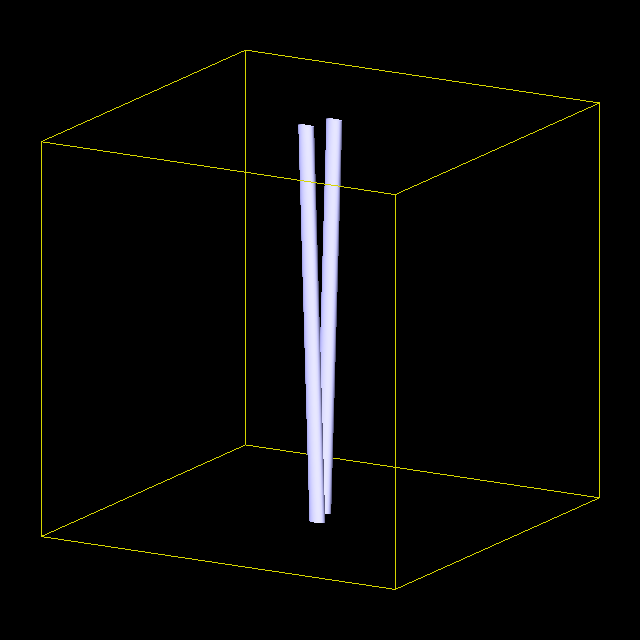}
  \includegraphics[width=3.3cm,bb=0 0 290 290]{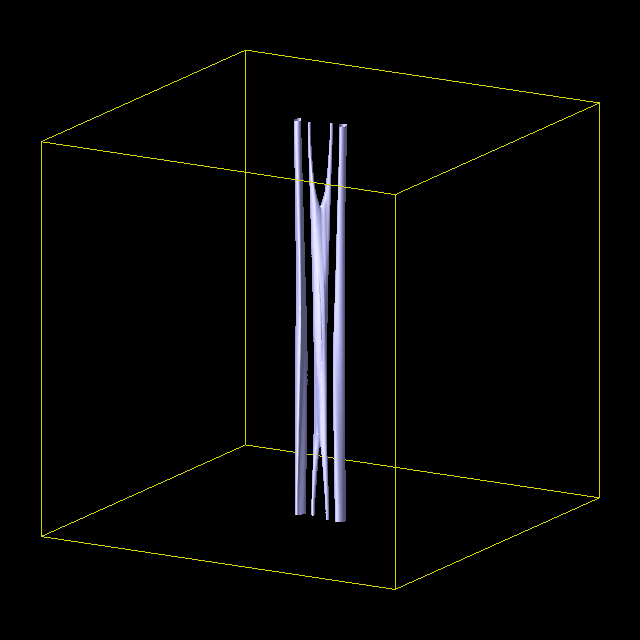}
  \includegraphics[width=3.3cm,bb=0 0 290 290]{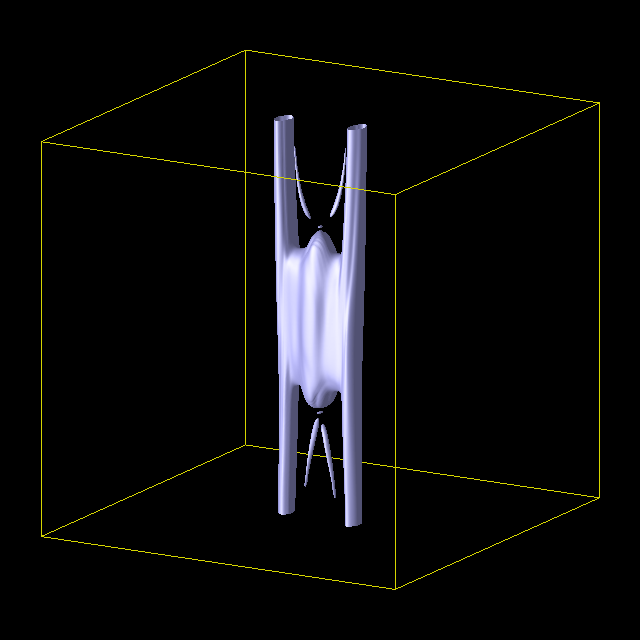}
  \includegraphics[width=3.3cm,bb=0 0 290 290]{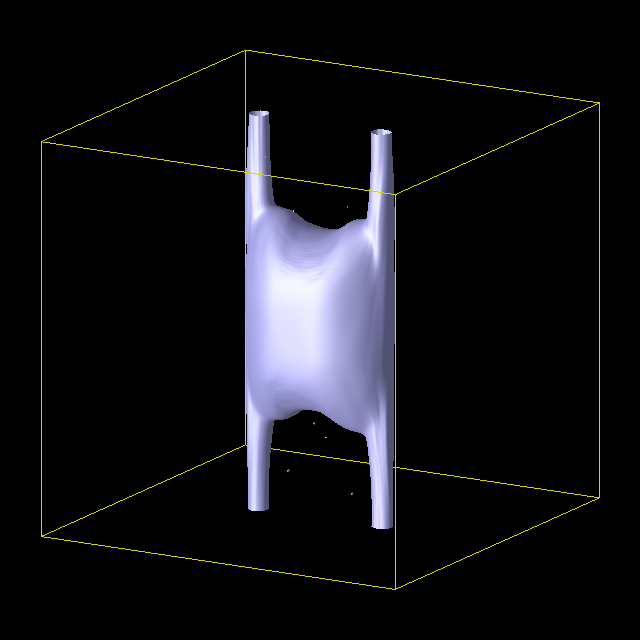}
  \includegraphics[width=3.3cm,bb=0 0 290 290]{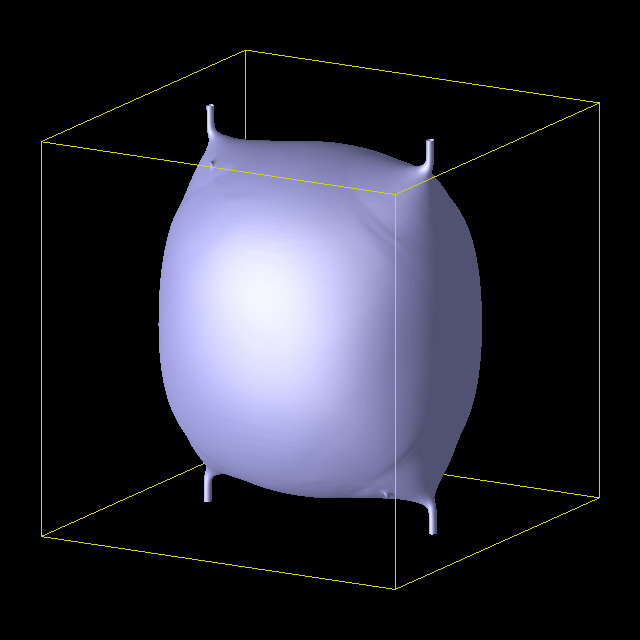}
}
\caption{[P3] Failed reconnection of parallel string pair with
 $(v,\alpha)=(0.94c,0.02\pi)$. See \cite{animation} for full
 animations.} 
\label{fig:3D_P3}
\end{figure}

\begin{figure}[!ht]
\centering{
  \includegraphics[width=3.3cm,bb=0 0 290 290]{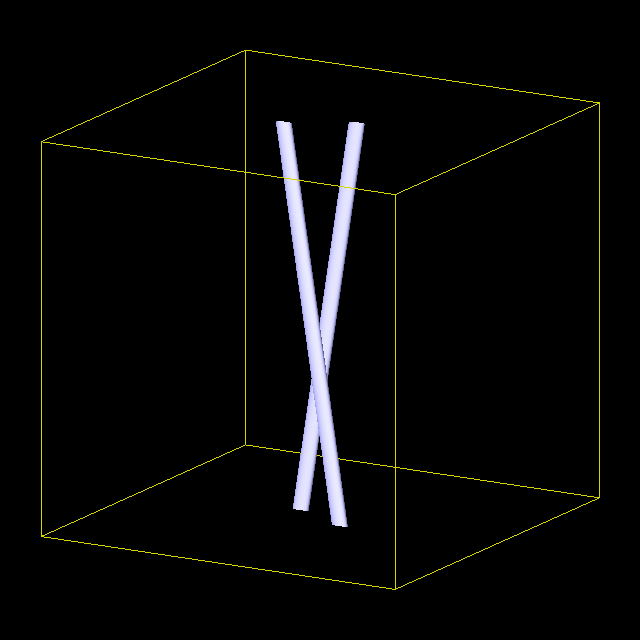}
  \includegraphics[width=3.3cm,bb=0 0 290 290]{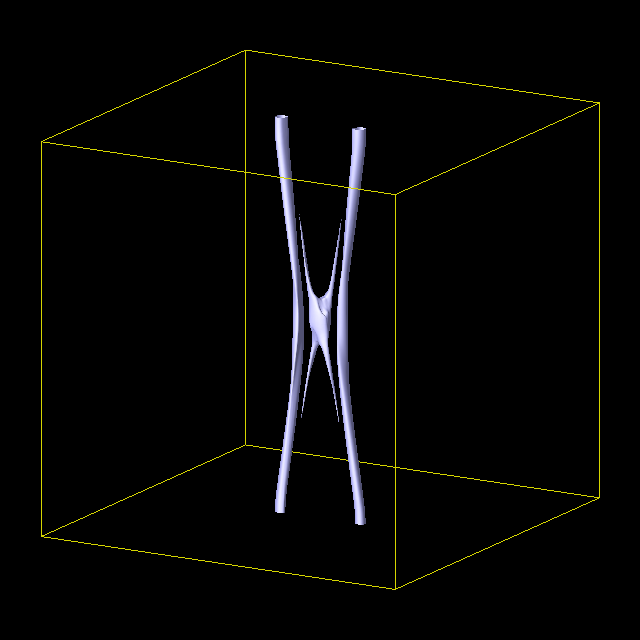}
  \includegraphics[width=3.3cm,bb=0 0 290 290]{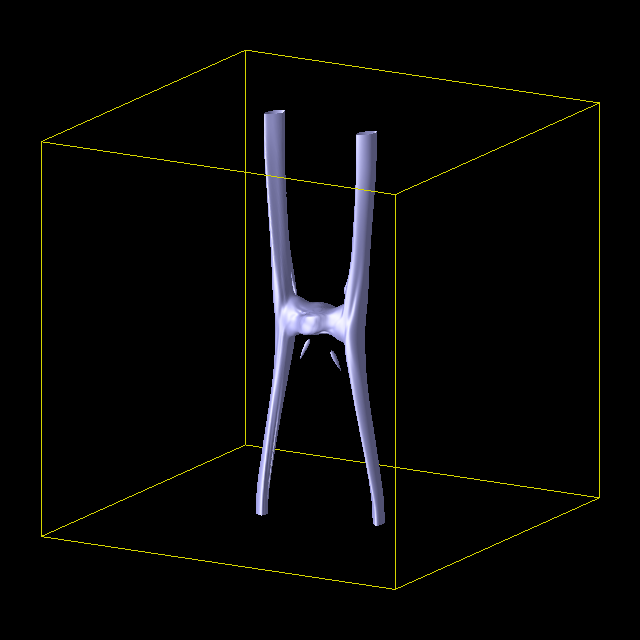}
  \includegraphics[width=3.3cm,bb=0 0 290 290]{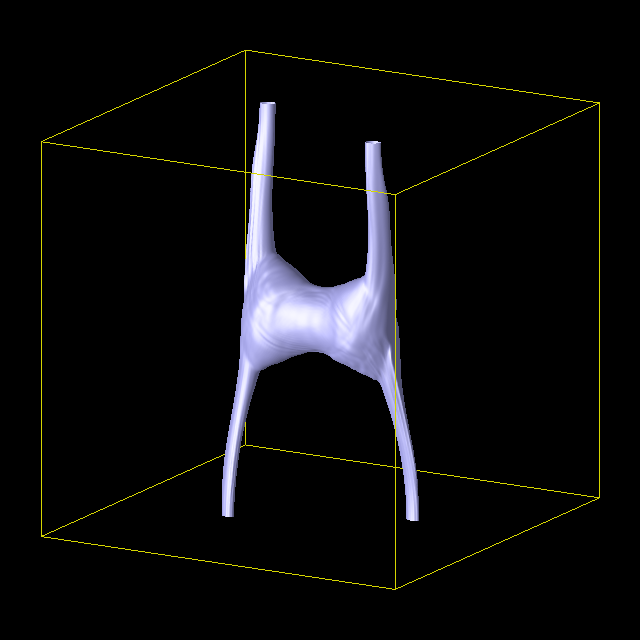}
  \includegraphics[width=3.3cm,bb=0 0 290 290]{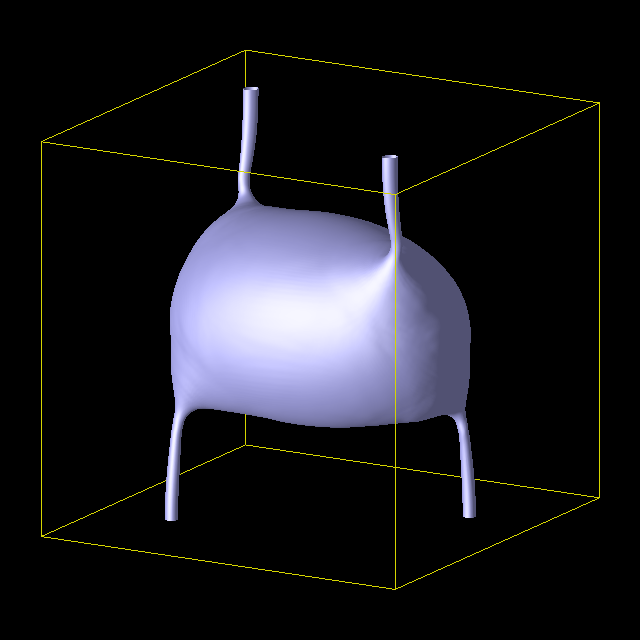}
}
\caption{[P4] Failed reconnection of parallel string pair with
 $(v,\alpha)=(0.95c,0.10\pi)$. See \cite{animation} for full
 animations.} 
\label{fig:3D_P4}
\end{figure}

\begin{figure}[!ht]
\centering{
  \includegraphics[width=3.3cm,bb=0 0 290 290]{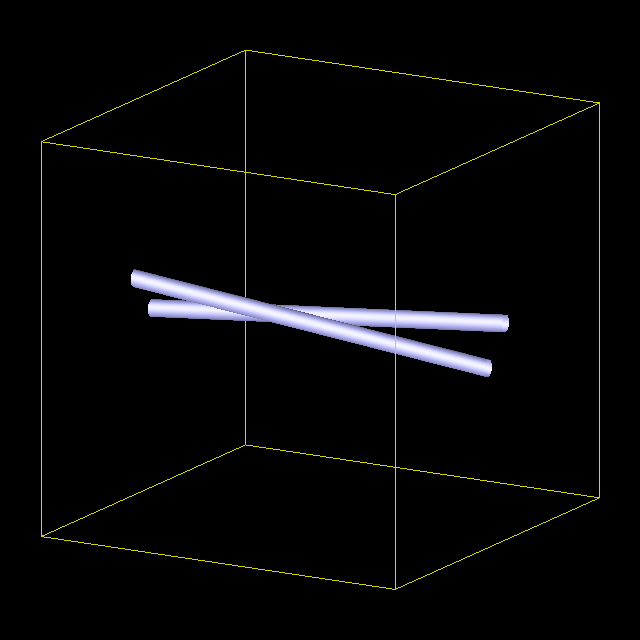}
  \includegraphics[width=3.3cm,bb=0 0 290 290]{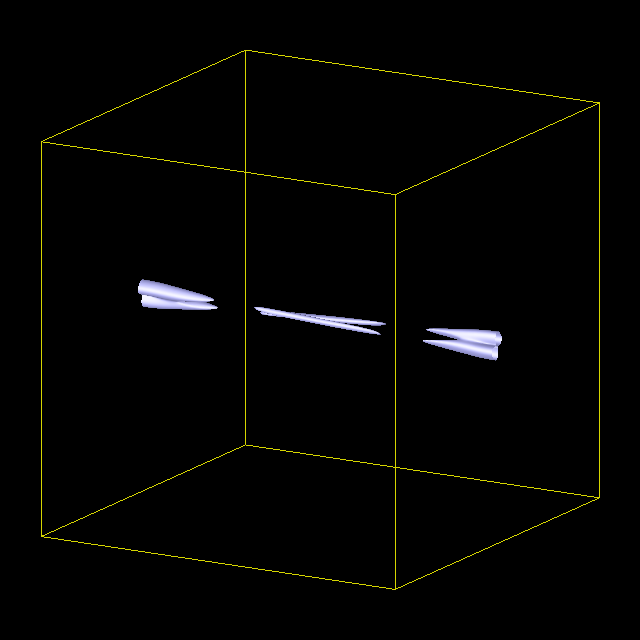}
  \includegraphics[width=3.3cm,bb=0 0 290 290]{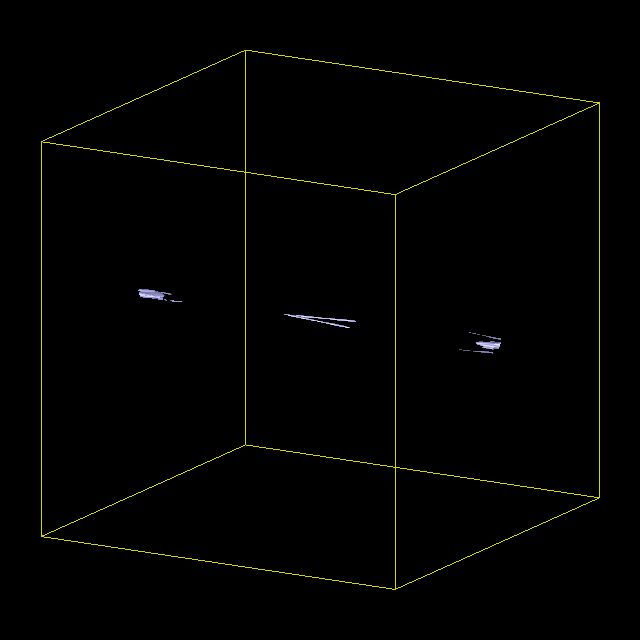}
  \includegraphics[width=3.3cm,bb=0 0 290 290]{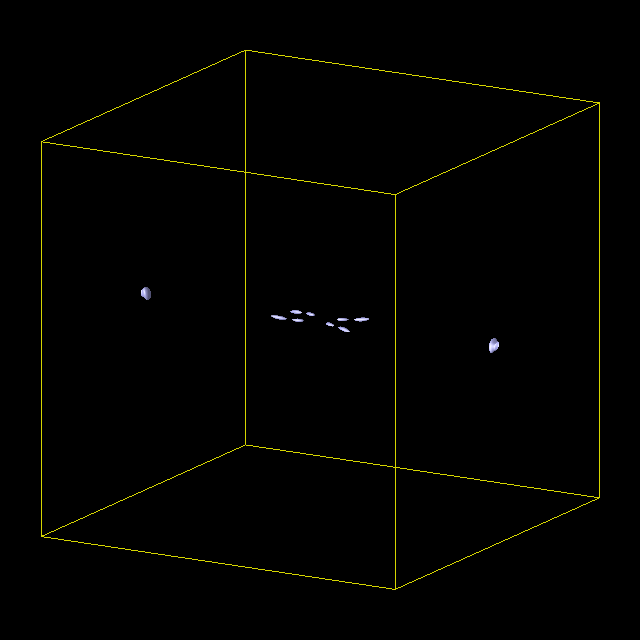}
  \includegraphics[width=3.3cm,bb=0 0 290 290]{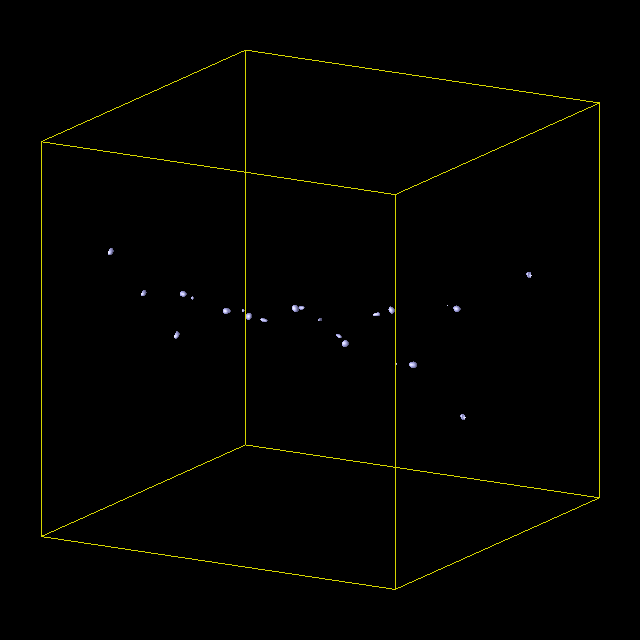}
}
\caption{[A1] Anti-parallel string pair with
 $(v,\alpha)=(0.62c,0.94\pi)$. Most of segments are annihilated. See
 \cite{animation} for full animations.} 
\label{fig:3D_A1}
\end{figure}

\begin{figure}[!ht]
\centering{
  \includegraphics[width=3.3cm,bb=0 0 290 290]{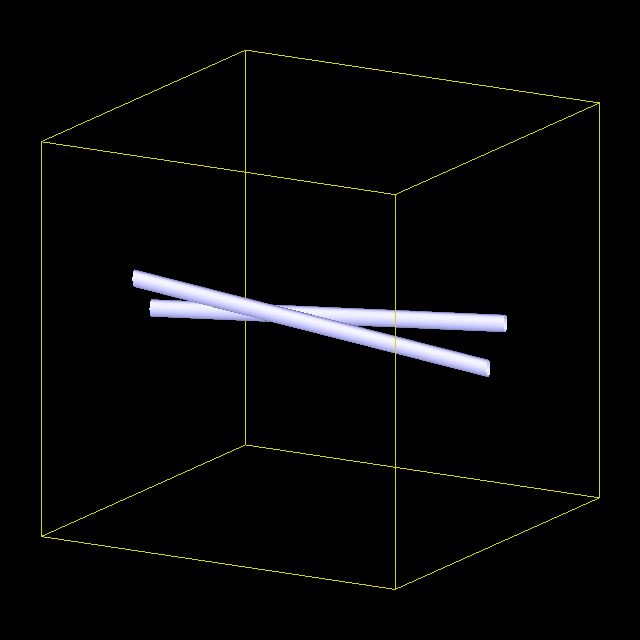}
  \includegraphics[width=3.3cm,bb=0 0 290 290]{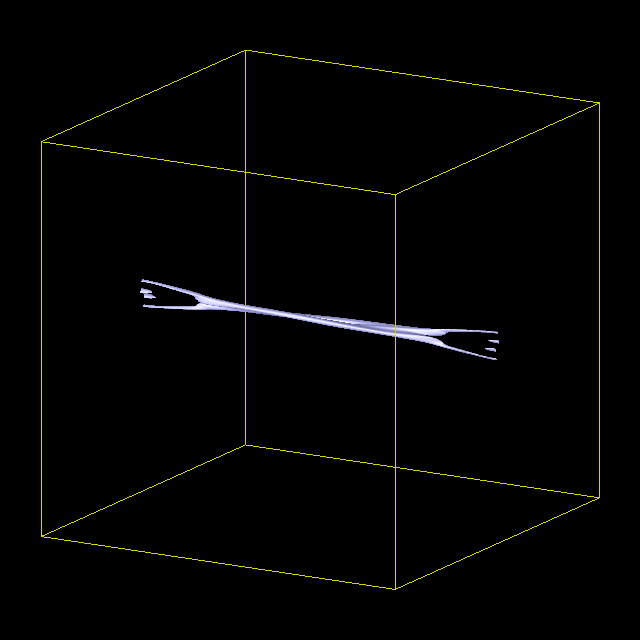}
  \includegraphics[width=3.3cm,bb=0 0 290 290]{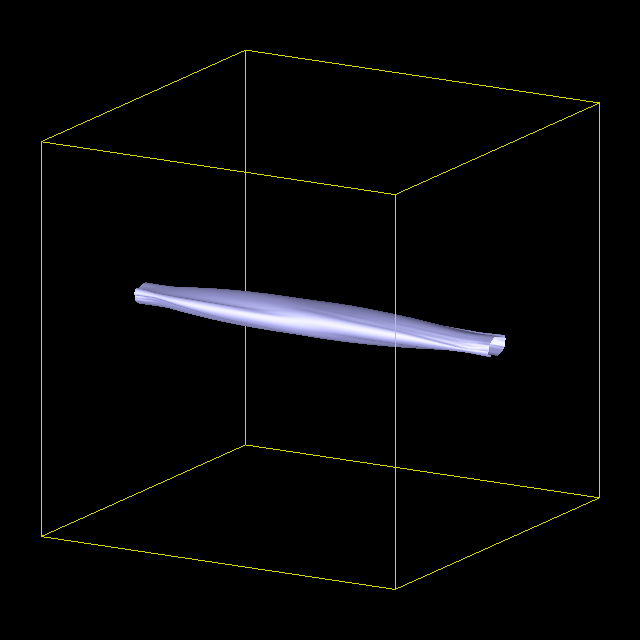}
  \includegraphics[width=3.3cm,bb=0 0 290 290]{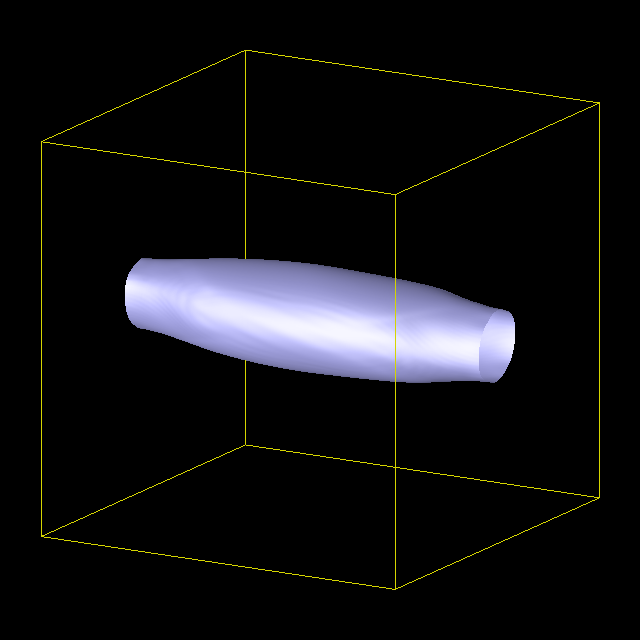}
  \includegraphics[width=3.3cm,bb=0 0 290 290]{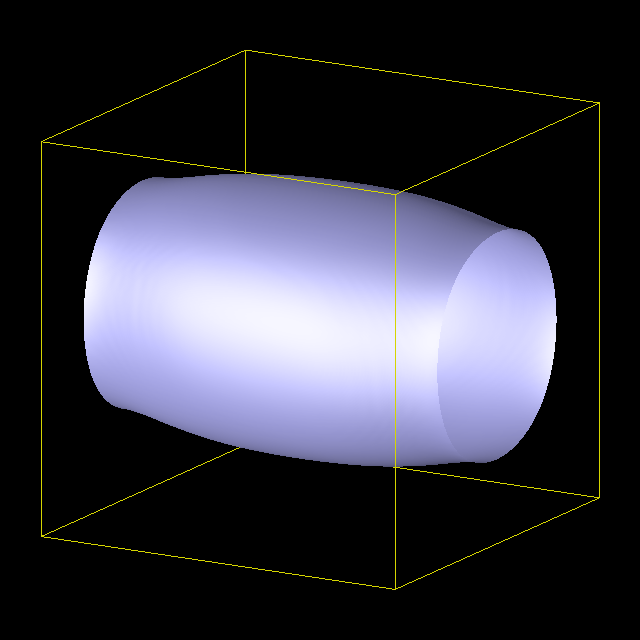}
}
\caption{[A2] Anti-parallel string pair with
 $(v,\alpha)=(0.90c,0.94\pi)$. After collision, the true vacuum region
 is created around the impact point, and it starts to grow
 exponentially. See \cite{animation} for full animations.}
\label{fig:3D_A2}
\end{figure}

\begin{figure}[!ht]
\centering{
  \includegraphics[width=4.0cm,bb=0 0 435 395]{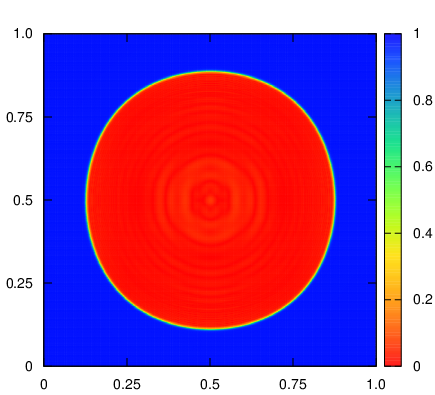}
  \includegraphics[width=4.0cm,bb=0 0 435 395]{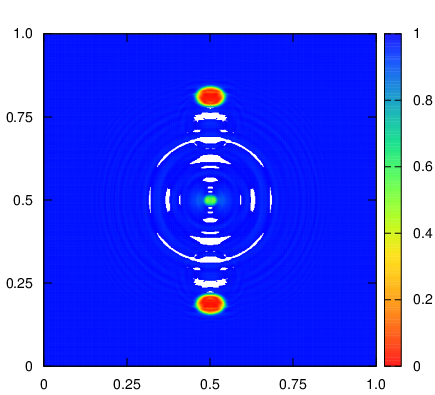}
  \includegraphics[width=4.0cm,bb=0 0 435 395]{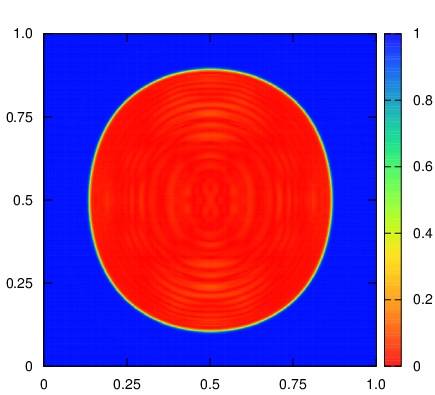}
  \includegraphics[width=4.0cm,bb=0 0 435 395]{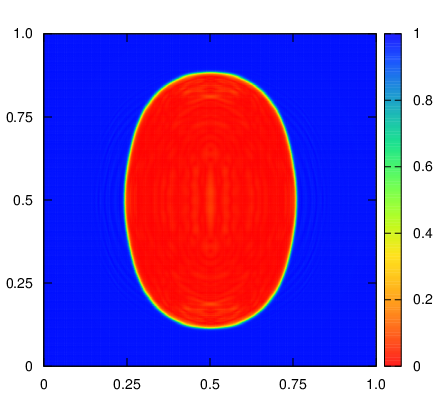}
}\\
\centering{
  \includegraphics[width=4.0cm,bb=0 0 435 395]{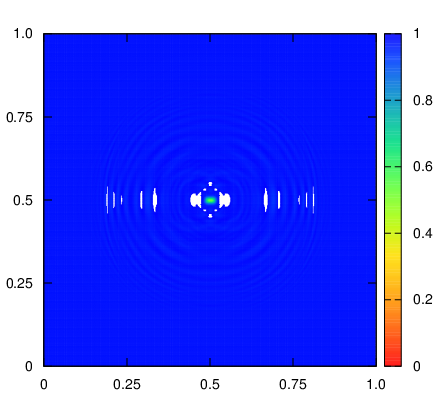}
  \includegraphics[width=4.0cm,bb=0 0 435 395]{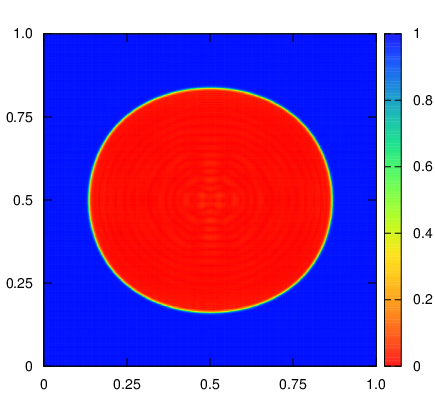}
}
\caption{The field configurations of [P1]-[P4] on $z=L/2$ (upper
 panels), and those of [A1], [A2] on $y=L/2 $(lower panels) 
at $t=t_f$. The colour contour represents 
$|X|/R_{\rm min}$. The white colour indicates the overshoot, 
$|X|>R_{\rm min}$.} 
\label{fig:2D}
\end{figure}

\begin{figure}[!ht]
\centering{
  \includegraphics[width=4.0cm,bb=0 0 435 395]{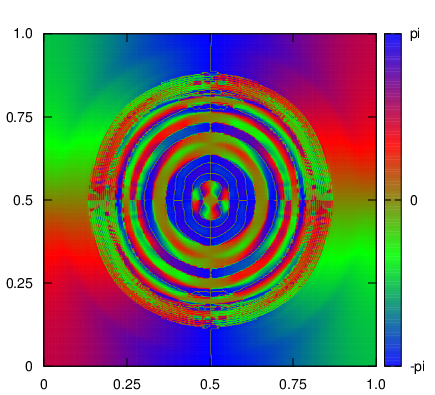}
  \includegraphics[width=4.0cm,bb=0 0 435 395]{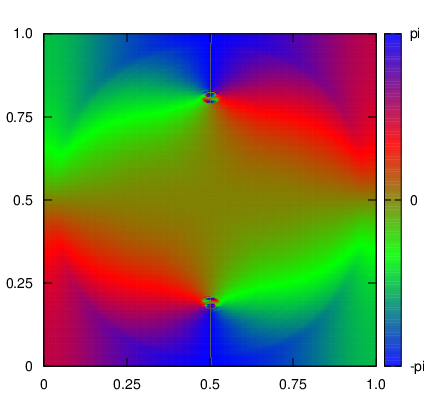}
  \includegraphics[width=4.0cm,bb=0 0 435 395]{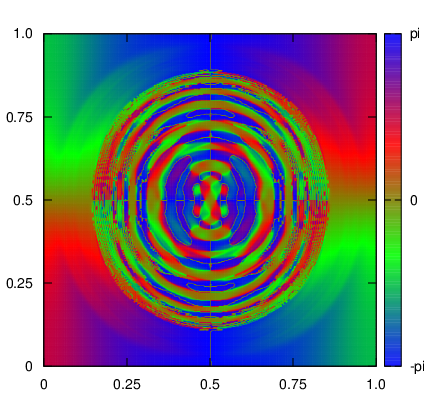}
  \includegraphics[width=4.0cm,bb=0 0 435 395]{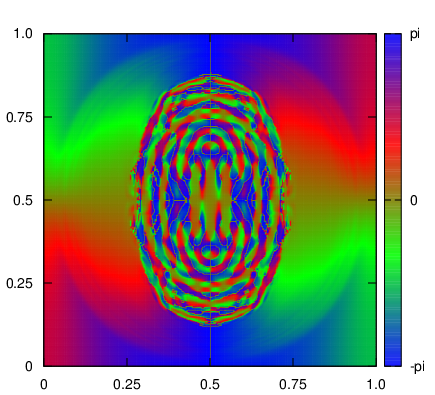}
}\\
\centering{
  \includegraphics[width=4.0cm,bb=0 0 435 395]{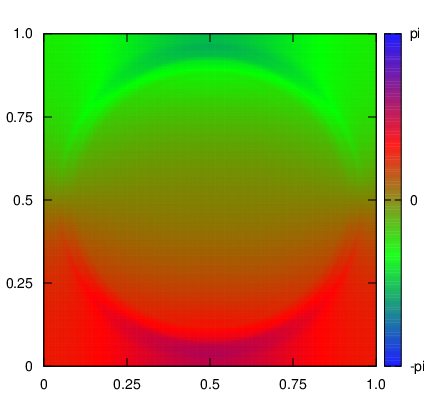}
  \includegraphics[width=4.0cm,bb=0 0 435 395]{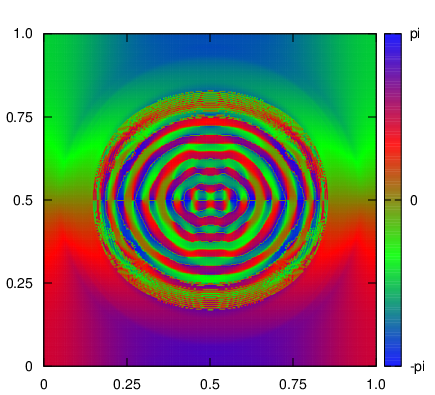}
}
\caption{The phases of [P1]-[P4] on $z=L/2$ (upper panels),
 and those of [A1], [A2] on $y=L/2$ (lower panels) at $t=t_f$. } 
\label{fig:2D_phase}
\end{figure}

Through the numerical simulations, we find that the collision processes
end up with either simply reconnecting and going away from each other,
or creating unstable objects with a higher winding number at the impact point.
Figs.~\ref{fig:3D_P1}-\ref{fig:3D_A2} show the snapshots of the
computational domain during simulations, the leftmost panel is at the 
initial time $t=0$, and the rightmost one at the final time $t=t_f$.
The surfaces in them represent $X=X_{\rm max}$ given in
Eq.~(\ref{eq:Xmaxmin}), and hence in the interior of them the field $X$
lies in the true vacuum. The first four figures,
Figs.~\ref{fig:3D_P1}-\ref{fig:3D_P4}, are the cases of parallel strings
with relatively small $\alpha$, and the remaining ones,
Figs.~\ref{fig:3D_A1} and \ref{fig:3D_A2}, are those of anti-parallel strings
with $\alpha \sim \pi$. 
Moreover, we plot the field configurations and those phases on the
surface at $z=L/2$ and $t=t_f$ in Figs.~\ref{fig:2D} and \ref{fig:2D_phase}.
In the unstable cases, P1, P3, P4 and A1, we find that the true vacuum
homogeneously spreads over the interior of the bubbles. As for the
phases, one can observe that the winding number is conserved.
For the parallel pairs, one can find that the total winding number is
$n=1+1=2$ (two sets of blue$\to$green$\to$red$\to$blue), if one follows
the trajectory around the bubble. In the interior, there seems to be a
large number of points around which the phase is rotated, although the
winding vanishes if one follows the trajectory around the impact point.
On the other hand, for the anti-parallel pairs, the total winding number
around the bubble becomes $n=1-1=0$
(green$\to$blue$\to$green$\to$red$\to$blue$\to$red$\to$green.)

%%%%%%%%%%%%%%%%%%%%%%%%%%%%%%%%%%%%%%%%%%%%%%%%%
%%%%%%%%%%%%%%%%%%%%%%%%%%%%%%%%%%%%%%%%%%%%%%%%%%

%%%%%%%%%%%%%%%%%%%%%%%%%%%%%%%%%%%%%%%%%%%%%%%%%%
%%%%%%%%%%%%%%%%%%%%%%%%%%%%%%%%%%%%%%%%%%%%%%%%%%

%%%%%%%%%%%%%%%%%%%%%%%%%%%%%%%%%%%%%%%%%%%%%%%%%%
%%%%%%%%%%%%%%%%%%%%%%%%%%%%%%%%%%%%%%%%%%%%%%%%%%

%%%%%%%%%%%%%%%%%%%%%%%%%%%%%%%%%%%%%%%%%%%%%%%%%%
%%%%%%%%%%%%%%%%%%%%%%%%%%%%%%%%%%%%%%%%%%%%%%%%%%

%%%%%%%%%%%%%%%%%%%%%%%%%%%%%%%%%%%%%%%%%%%%%%%%%%
%%%%%%%%%%%%%%%%%%%%%%%%%%%%%%%%%%%%%%%%%%%%%%%%%%

%%%%%%%%%%%%%%%%%%%%%%%%%%%%%%%%%%%%%%%%%%%%%%%%%%
%%%%%%%%%%%%%%%%%%%%%%%%%%%%%%%%%%%%%%%%%%%%%%%%%%

%%%%%%%%%%%%%%%%%%%%%%%%%%%%%%%%%%%%%%%%%%%%%%%%%%
%%%%%%%%%%%%%%%%%%%%%%%%%%%%%%%%%%%%%%%%%%%%%%%%%%

%%%%%%%%%%%%%%%%%%%%%%%%%%%%%%%%%%%%%%%%%%%%%%%%%
%%%%%%%%%%%%%%%%%%%%%%%%%%%%%%%%%%%%%%%%%%%%%%%%%

\begin{figure}[!ht]
\centering{
  \includegraphics[width=12cm,bb=0 50 792 550]{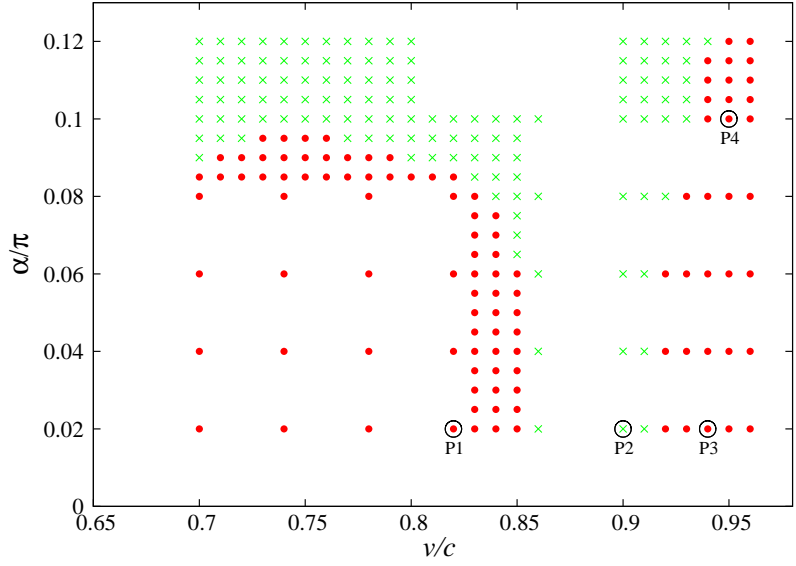}
}
\caption{The parameter region for stable/unstable collision process for
 string-string collision.}
\label{fig:ss}
\end{figure}

\begin{figure}[!ht]
\centering{
  \includegraphics[width=12cm,bb=0 50 792 550]{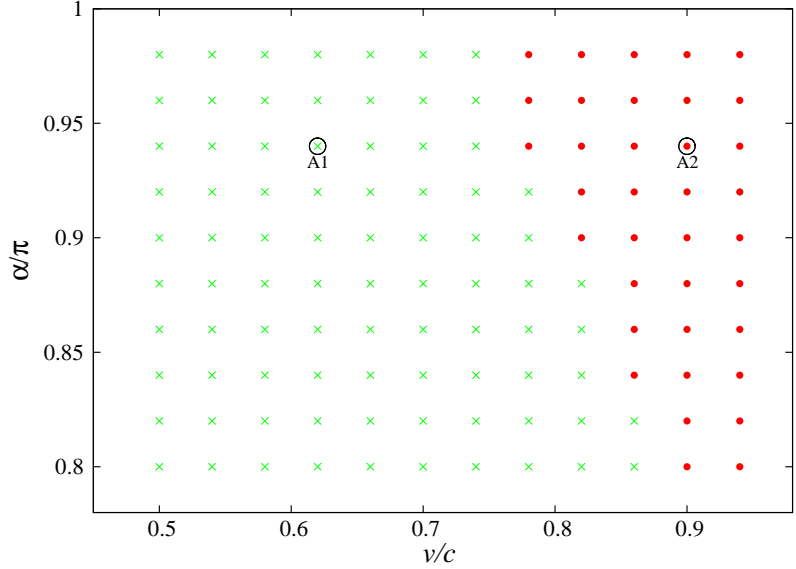}
}
\caption{The parameter region for stable/unstable collision process for
 string-antistring collision.}
\label{fig:sa}
\end{figure}

The (in)stability of a resultant string after collision depends on
the velocity of two cylinders, $v$, and the collision angle between
them, $\alpha$. We surveyed the parameter space $(\alpha,v)$ to check
the stability.
Fig.~\ref{fig:ss} shows the stability of parallel string pairs,
and Fig.\ref{fig:sa} that of anti-parallel pairs
where $\alpha$ is close to $\pi$.
In order to systematically judge the resultant stability, we calculate
two quantities and set a criterion for each. One is the growth rate of
the volume of the true vacuum region during simulation, 
%%%%%%%%%%%%%%%%%%%%%%%%%%%%%%%%%%%%%%%%%%%%%%%%%%%%%%%%%%%%%%%%%%%%
%
\begin{equation}
 \kappa_1 = \frac{\mathcal{V}_{\rm true}(t=t_f)}{\mathcal{V}_{\rm true}(t=0)},
\end{equation}
%
%%%%%%%%%%%%%%%%%%%%%%%%%%%%%%%%%%%%%%%%%%%%%%%%%%%%%%%%%%%%%%%%%%%%
and the other is the ratio of time when the true vacuum region grows to
the total simulation time, 
%%%%%%%%%%%%%%%%%%%%%%%%%%%%%%%%%%%%%%%%%%%%%%%%%%%%%%%%%%%%%%%%%%%%
%
\begin{equation}
 \kappa_2 = \frac{t_{\rm grow}}{t_f},\qquad
 t_{\rm grow} = \Delta t \times \sharp\left\{ t_m \Bigg| \left.\frac{d\mathcal{V}_{\rm true}}{dt}\right|_{t=t_m}>0\right\},
\end{equation}
%
%%%%%%%%%%%%%%%%%%%%%%%%%%%%%%%%%%%%%%%%%%%%%%%%%%%%%%%%%%%%%%%%%%%%
where $\sharp$ denotes the number of elements, $\Delta t$ is the time
interval, and $t_m$ represents the discrete time, $t_m = m\Delta t$.
If and only if both $\kappa_1 > \kappa_{1c}$ and 
$\kappa_2 > \kappa_{2c}$ are satisfied, we decide that the resultant
strings become unstable, and thus the true vacuum region grows
exponentially. We set $\kappa_{1c} = 10$ and $\kappa_{2c} = 0.8$. 
The second quantity, $\kappa_2$, measures how monotonic the
instability grows. After all, however, most of (in)stabilities can be
determined by the first one, $\kappa_1$.

In Figs.~\ref{fig:ss} and \ref{fig:sa}, we plot red filled circles
for unstable pairs, and green crosses for stable pairs.
We find that the slow collision with small $\alpha$ results in the
failure of the reconnection and thus being unstable, and quite
high-speed collision can also make the system unstable in both cases
with small (parallel pair) and large $\alpha$ (anti-parallel pair).
In these figures, we also plot black circles labelled as P1,2,3,4 and
A1,2. The corresponding figures were already shown in
Figs.~\ref{fig:3D_P1}-\ref{fig:3D_A2}, which are the 
typical examples of several regions leading to stable and
unstable processes in Figs.~\ref{fig:ss} and \ref{fig:sa}.

We explain in detail the mechanism of stable/unstable reconnection
processes using Figs.~\ref{fig:3D_P1}-\ref{fig:3D_A2}.
Firstly, for the parallel pairs, our findings are followings.
\begin{itemize}
\item [P1] (relatively slow collision) : Two strings merge at the instance
      of impact and cannot be separated from each other because of a small
      collision velocity. Then the effective volume of the merged
      strings around the impact point exceeds the critical volume
      mentioned in Sec.\ref{sec:ana}, thus making the system unstable.
\item [P2] (moderate-speed collision) : Two strings can safely
      reconnect with each other. As shown in the middle panel and its
      right neighbour, a small bubble is created at the impact points of
      the strings. However, since the reconnected strings rapidly go
      away from the bubble and the volume of the bubble is relatively
      small, the bubble finally shrinks. As a result, the system becomes stable.
\item [P3] (high-speed collision) : Because of the large collision velocity, 
      the kinetic energy induced to the bubble at the impact point
      is large enough to inflate the bubble until its volume exceeds the
      critical one. Finally the system becomes unstable.
\item [P4] (large angle high-speed collision ) : Since the collision
      angle is large, the initial impact of strings cannot create a larger
      bubble at the impact point than the critical volume. However,
      kinetic energy induced to the bubble is too large, and then the
      bubble can rapidly expand (the third panel). At the same time, the
      reconnected strings also rapidly go away from the impact point. As
      a result, the bubble is about to break up (forth panel). Finally,
      the expansion rate of the bubble overcomes (last panel).
\end{itemize}

As for the anti-parallel pairs, they annihilate with each other in most
of cases owing to the cancellation of their winding numbers [A1]. 
It is, however, surprising that the high-speed collision makes the
system unstable. Actually, after the collision, a bubble with no
windings is excited at the impact point, and it grows exponentially. 

%%%%%%%%%%%%%%%%%%%%%%%%%%%%%%%%%%%%%%%%%%%%%%%%%%

%%%%%%%%%%%%%%%%%%%%%%%%%%%%%%%%%%%%%%%%%%%%%%%%%%

%%%%%%%%%%%%%%%%%%%%%%%%%%%%%%%%%%%%%%%%%%%%%%%%%%%%%%%%%%%%%%%%%%%%%%%%%%%%%%%
%%%%%%%%%%%%%%%%%%%%%%%%%%%%%%%%%%%%%%%%%%%%%%%%%%%%%%%%%%%%%%%%%%%%%%%%%%%%%%%
\section{Conclusion}
\label{sec:conc}
%%%%%%%%%%%%%%%%%%%%%%%%%%%%%%%%%%%%%%%%%%%%%%%%%%%%%%%%%%%%%%%%%%%%%%%%%%%%%%%
%%%%%%%%%%%%%%%%%%%%%%%%%%%%%%%%%%%%%%%%%%%%%%%%%%%%%%%%%%%%%%%%%%%%%%%%%%%%%%%

We have studied the dynamical (in)stabilities of metastable strings
after their collision in the model where the potential of a complex scalar
field has a false vacuum state corresponding to a SUSY breaking
vacuum in realistic SUSY models.

Before performing numerical simulations for the collision, we
analytically investigated in a simplified model the thickness of the
domain wall constituting the surface of strings, and the existence of
static solutions with large winding numbers from the viewpoint of energetics. 
As a result, we found that the thickness of wall is determined
from the shape of given potential, being independent of the
winding number of strings,
and there exists the upper bound of winding number for the stable
solutions. Furthermore, in the same manner, we investigated 
the collision process of two metastable strings with some
approximations neglecting the details of dynamics. 
Then we found that there is the critical volume of the true vacuum
region in a string as indicated in Eq.~(\ref{eq:critical_radius}) so
that the instability grows, and thus in some 
cases with a small collision angle, the reconnection is failed.

After that, we performed three-dimensional field-theoretic
simulations of two colliding metastable strings. The initial condition
is given as the superposition of two Lorentz-boosted strings obtained in
another numerical way. 
We surveyed the (in)stabilities of the collision processes on the
parameter space $(v,\alpha)$, where $v$ is the velocity of strings and
$\alpha$ is the collision angle. 

Consequently, we found that the instability cannot always be observed
for the string-string pairs. We fixed the winding number of
strings as $n=1$ and tuned the parameters controlling the shape of
potential, $\epsilon,\zeta$ and $\delta$, so that the static solution
with $n=2$ does not exist. Nevertheless the strings with most of
combinations $(v,\alpha)$ lead to successful reconnection, or, in other words,
the parameter region in the $(v,\alpha)$ space where the instability
grows is highly restricted. 

We briefly explain the unstable cases. For small $\alpha$, 
we confirmed that the reconnection is failed owing to the rapid expansion of the
overlap region just after the collision, shown in Fig.~\ref{fig:3D_P1} [P1] and  
Fig.~\ref{fig:3D_P3} [P3], as discussed in the static analysis mentioned above.
In addition to this case, we found that the kinetic
energy injected into the impact point can leads to failure reconnection.
In fact, although the strings with a large velocity seem to succeed in
reconnecting safely, the whole system becomes eventually unstable owing to the
rapid growth of a true vacuum bubble created at the impact
point after the collision, as shown in Fig.~\ref{fig:3D_P3} [P3] and Fig.~\ref{fig:3D_P4} [P4].
Surprisingly, this is also the case for the anti-parallel pair, as shown
in Fig.~\ref{fig:3D_A2} [A2], although one would envisage that they can
pair-annihilate. This result would be
explained by the shorter time scale of the expansion of the zero-winding
bubble created at the impact point than that of pair-annihilation. 

At first, we had a naive expectation such that
the instability would grow because of the temporal formation of unstable $n=2$
strings at the moment of impact, and the anti-parallel pairs always
annihilate after collision. Our numerical studies, however,
clarified that this is not true story. Instead of the winding
number iteself, it is confirmed that 
the volume of true vacuum at the impact point and the collision veclocity are
responsible for the (in)stability of the colliding strings.
Particularly, it should be stressed that the numerical studies presented
here are crucial to find the latter fact, dependence on the velocity or
kinetic energy of strings.

%\hir{
%Our results have significant implications on model building for
%realistic SUSY breaking. The SUSY breaking models often lead to
%the formation of tube-like strings, but such strings should be stable
%even by collisions, which we studied here. That would constrain the
%viable prameter space of SUSY breaking models and also cosmological
%history.
%}

Our numerical study has shown the relatively complicated parameter
dependence on the stability of tube-like strings against collisions, and
will help to constrain the viable parameter 
space on concrete SUSY breaking models with spontaneous $U(1)_R$ symmetry 
breaking and also their cosmological history.

\begin{acknowledgments}
T.~H. was supported by JSPS Grant-in-Aid for Young Scientists (B)
No.23740186, and by MEXT HPCI Strategic Program. The work of M.~E. is supported by
Grant-in-Aid for Scientific Research from the Ministry of Education,
 Culture, Sports, Science and Technology, Japan (No. 23740226) 
and Japan Society for the Promotion of Science (JSPS) and Academy of
Sciences of the Czech Republic (ASCR) under the Japan - Czech Republic
Research Cooperative Program. T.~K. is supported in part by the
Grant-in-Aid for the Global COE Program "The Next Generation of Physics,
Spun from Universality and Emergence" and the JSPS Grant-in-Aid for
Scientific Research (A) No. 22244030 from the Ministry of Education,
Culture,Sports, Science and  Technology of Japan. Y.~O.'s research is
supported by The Hakubi Center for Advanced Research, Kyoto University. 
\end{acknowledgments}

%%%%%%%%%%%%%%%%%%%%%%%%%%%%%%%%%%%%%%%%%%%%%%%%%%%%%%%%%%%%%%%%%%%%%%%%%%%%%%%
%%%%%%%%%%%%%%%%%%%%%%%%%%%%%%%%%%%%%%%%%%%%%%%%%%%%%%%%%%%%%%%%%%%%%%%%%%%%%%%
%%%%%%%%%%%%%%%%%%%%%%%%%%%%%%%%%%%%%%%%%%%%%%%%%%%%%%%%%%%%%%%%%%%%%%%%%%%%%%%
%%%%%%%%%%%%%%%%%%%%%%%%%%%%%%%%%%%%%%%%%%%%%%%%%%%%%%%%%%%%%%%%%%%%%%%%%%%%%%%

\appendix
%%%%%%%%%%%%%%%%%%%%%%%%%%%%%%%%%%%%%%%%%%%%%%%%%%%%%%%%%%%%%%%%%%%%%%%%%%%%%%%
%%%%%%%%%%%%%%%%%%%%%%%%%%%%%%%%%%%%%%%%%%%%%%%%%%%%%%%%%%%%%%%%%%%%%%%%%%%%%%%
\section{numerical schemes}
\label{appsec:numerics}
%%%%%%%%%%%%%%%%%%%%%%%%%%%%%%%%%%%%%%%%%%%%%%%%%%%%%%%%%%%%%%%%%%%%%%%%%%%%%%%
%%%%%%%%%%%%%%%%%%%%%%%%%%%%%%%%%%%%%%%%%%%%%%%%%%%%%%%%%%%%%%%%%%%%%%%%%%%%%%%

To find the static configuration of an axially symmetric metatable
string, we solve Eq.~(\ref{eq:R}) with the succesive over-relaxation
method. 

First of all, we discretise the coordinate such as $x_j=j\Delta x$ for
$j=0,1,\ldots,N$, and represent the discretised $R(x)$ as a vector
$R_j\equiv R(x_j)$. The spatial derivatives in the left-hand side of
Eq.~(\ref{eq:R}) is replaced by the corresponding 2nd-order finite
differences,
%%%%%%%%%%%%%%%%%%%%%%%%%%%%%%%%%%%%%%%%%%%%%%%%%%%%%%%%%%%%%%%%%%%%
%
\begin{equation}
 \left.\frac{d^2R}{dx^2}\right|_{x=x_j} \approx \frac{R_{j+1}-2R_j+R_{j-1}}{\Delta x^2}, \qquad
 \left.\frac{dR}{dx}\right|_{x=x_j} \approx \frac{R_{j+1}-R_{j-1}}{2\Delta x}.
\end{equation}
%
%%%%%%%%%%%%%%%%%%%%%%%%%%%%%%%%%%%%%%%%%%%%%%%%%%%%%%%%%%%%%%%%%%%%
We denote the other terms depending on $R(x)$ in Eq.~(\ref{eq:R}) by $S[R(x)]$ and those
evaluated at $x=x_j$ by $S_j[R]\equiv S[R(x_j)]$.
Then the equation to be solved becomes
%%%%%%%%%%%%%%%%%%%%%%%%%%%%%%%%%%%%%%%%%%%%%%%%%%%%%%%%%%%%%%%%%%%%
%
\begin{equation}
 \frac{R_{j+1}-2R_j+R_{j-1}}{\Delta x^2} + \frac{R_{j+1}-R_{j-1}}{2x_j\Delta x}
 = S_j[R].
\end{equation}
%
%%%%%%%%%%%%%%%%%%%%%%%%%%%%%%%%%%%%%%%%%%%%%%%%%%%%%%%%%%%%%%%%%%%%
Solving this equations with respect to
$R_j$ in the first term, we find trial values of $R_j$ denoted by $R_j^{*}$,
%%%%%%%%%%%%%%%%%%%%%%%%%%%%%%%%%%%%%%%%%%%%%%%%%%%%%%%%%%%%%%%%%%%%
%
\begin{equation}
 R_j^{*} = 
 \frac{R_{j+1}+R_{j-1}}{2}-\frac{\Delta x^2}{2}\left(S_j[R] - \frac{R_{j+1}-R_{j-1}}{2x_j\Delta x}\right).
\end{equation}
%
%%%%%%%%%%%%%%%%%%%%%%%%%%%%%%%%%%%%%%%%%%%%%%%%%%%%%%%%%%%%%%%%%%%%
Notice that $R_j$ in $S_j[R]$ is not $R^*_j$.
On each stage of iterations, we update $R_j$ so that
%%%%%%%%%%%%%%%%%%%%%%%%%%%%%%%%%%%%%%%%%%%%%%%%%%%%%%%%%%%%%%%%%%%%
%
\begin{equation}
 R_j^{(n+1)} = \omega R_j^{*} + (1-\omega)R_j^{(n)}.
\end{equation}
%
%%%%%%%%%%%%%%%%%%%%%%%%%%%%%%%%%%%%%%%%%%%%%%%%%%%%%%%%%%%%%%%%%%%%
This scheme with $\omega=1$ is equivalent to the Gauss-Seidel method.
When we solve Eq.~(\ref{eq:R}) for $\zeta=4.0$ in
Sec.~\ref{sec:model}, we set $\omega=1, 0.5$ and $0.3$ for $n=1,2,3$,
respectively. For $\zeta=2.2$ in Sec.~\ref{sec:setup}, we set $\omega=1$.

\end{document}